\documentclass[a4paper,11pt]{article}
\pdfoutput=1 

\usepackage{jheppub, dsfont, amssymb} 

\usepackage[T1]{fontenc} 
\usepackage{multirow}
\usepackage{xcolor}
\usepackage{graphics, graphicx}

\newcommand{\Ref}[1]{(\ref{#1})}

\newcommand{\be}{\begin{equation}}
\newcommand{\ee}{\end{equation}}
\newcommand{\bea}{\begin{eqnarray}}
\newcommand{\eea}{\end{eqnarray}}

\newcommand{\bit}{\begin{itemize}}
\newcommand{\eit}{\end{itemize}}

\newcommand{\tr}{{\rm Tr}}

\def\p{\partial}

\newcommand{\ii}{\mathrm{i}}
\newcommand{\ex}{\mathrm{e}}
\newcommand{\dd}{\mathrm{d}}

\title{\boldmath Group field theory for quantum gravity minimally coupled to a scalar field}

\author[a,b]{Yang Li}
\author[b]{Daniele Oriti}
\author[b]{and Mingyi Zhang}

\affiliation[a]{Department of Physics, Beijing Normal University,\\Beijing 100875, China}
\affiliation[b]{Max-Planck-Institut f\"ur Gravitationsphysik (Albert-Einstein-Institut),\\Am M\"uhlenberg 1, 14476 Golm, Germany}

\emailAdd{liyang12@mail.bnu.edu.cn}
\emailAdd{daniele.oriti@aei.mpg.de}
\emailAdd{mingyi.zhang@aei.mpg.de}

\abstract{We construct a group field theory model for quantum gravity minimally coupled to relativistic scalar fields, defining as well a corresponding discrete gravity path integral (and, implicitly, a coupled spin foam model) in its Feynman expansion. We also analyze a number of variations of the same model, the corresponding discrete gravity path integrals, its generalization to the coupling of multiple scalar fields and discuss its possible applications to the extraction of effective cosmological dynamics from the full quantum gravity formalism, in the context of group field theory condensate cosmology.
}

\begin{document}
\maketitle
\flushbottom

\section{Introduction}\label{sec:intro}
\noindent In the group field theory (GFT) framework for quantum gravity \cite{Oriti:2011jm,Krajewski:2012aw,Oriti:2014uga,Oriti2015}, the fundamental degrees of freedom of quantum space-time, the GFT quanta, carry algebraic data which endow them with an interpretation in terms of discrete quantum geometries. They can be interpreted, in fact, as quantized tetrahedra which, via appropriate conditions imposed on the quantum states of the theory, can be \lq glued\rq to one another to form extended d-dimensional triangulations.  The same algebraic data characterize the spin network states in loop quantum gravity \cite{Ashtekar2004,Rovelli2004,Thiemann2008,MUXIN2007}, which indeed can be seen as associated to graphs dual to the triangulations formed by GFT quanta and correspond to an expansion of GFT states into an eigenbasis of specific discrete geometric operators (areas of triangles and volumes of tetrahedra). From this perspective, the GFT framework can be understood as a 2nd quantized, field-theoretic formulation of loop quantum gravity \cite{Oriti:2013aqa}. The correspondence between discrete quantum geometric structures and spin networks carries through at the dynamical level. The GFT dynamics of these degrees of freedom is treated via standard QFT methods. In particular, the partition function and quantum n-point functions of any given GFT model can be computed in perturbative expansion with respect to its coupling constants. The GFT Feynman amplitudes, associated to diagrams encoding the history of interacting GFT quanta and dual to (d+1)-dimensional triangulations, take the form of discrete path integral (DPI) of gravity discretized on such triangulations. The full quantum dynamics involves therefore naturally a sum over triangulations of all topologies as a well as a sum over discretized quantum geometries. The same Feynman amplitudes have a dual re-writing as spin foam models \cite{Alexandrov:2011ab,Perez:2012wv}, which are another (covariant) way of defining the dynamics of loop quantum gravity spin networks (and generalizations thereof).

\

\noindent The formulation of quantum gravity in terms of discrete path integrals has of course a long tradition and it is still vigorously pushed (see for example \cite{Ambjorn:2012jv}). What GFTs add is, on the one hand, the close relation with loop quantum gravity and the many insights and mathematical results obtained there and, on the other hand, a new perspective and powerful analytical tools due to the field-theoretical embedding of the same discrete gravity path integrals. These additional tools are especially useful for tackling the two related problems of defining the continuum limit of the theory and extracting effective continuum physics from it. Indeed, GFT quanta are seen as building blocks of the space-time, and the continuum space-time we use at the macroscopic, effective level is supposed to emerge from their collective dynamics in a regime where a large (or infinite) number of them is taken into account. For both issues, in fact, field-theoretic GFT techniques have proven very promising. The continuum limit of the discrete quantum geometric dynamics has been defined, at least for simple GFT models both using constructive techniques (see \cite{Rivasseau:2016rgt} for an introduction to these techniques), which correspond to a direct re-summation of the sum over triangulations weighted by the GFT Feynman amplitudes, and by means of a functional renormalization group approach to the definition of the GFT partition function, bypassing its formulation in terms of discrete path integrals altogether. For a review of the GFT renormalization programme, see \cite{Carrozza:2016vsq}, while for recent works in the functional formulation, see \cite{Benedetti:2014qsa, Geloun:2016qyb, Carrozza:2016tih}. One main strategy that has been followed to extract effective continuum physics from GFTs, in addition to those focusing exclusively on spin network and spin foam structures or on the sum over triangulations in related formalisms, has taken also advantage of the field theory aspects of the  framework and was based on GFT condensate states and their (mean field) dynamics. See \cite{Gielen:2016dss,Oriti:2016acw} for reviews and \cite{Oriti:2015rwa,Oriti:2016qtz} for some recent results.

\

\noindent The goal of extracting effective continuum physics from GFT models (and related formalisms) requires one more crucial ingredient: the inclusion in the same models of matter degrees of freedom (including other interaction fields).
One reason is obvious: there is plenty of other matter in the universe and the physics we are interested in understanding from a quantum gravity perspective relates (also) to matter fields and their interactions. The second reason is less obvious but equally fundamental: in a fully background independent and diffeomorphism invariant context like the one defined by GFTs and related formalisms, matter fields are the most convenient way to define physical reference frames, spatial and temporal directions to use to label the dynamics and the interactions of other degrees of freedom. This is the relation approach to the definition of physical observables in quantum gravity \cite{Rovelli:1990ph,Brown:1994py,Dittrich:2005kc}. In particular, the use of a free massless scalar field allows a simple definition of temporal evolution while maintaining manifest diffeomorphism invariance and it is the standard choice in quantum cosmology. It is therefore also the simplest choice to adopt in a full quantum gravity framework, when aiming at the extraction of effective cosmological dynamics from it. GFT condensate cosmology is in fact a case at hand \cite{Oriti:2016qtz}.

\

\noindent With the same basic motivations, in this paper we tackle the problem of coupling scalar matter fields in a GFT, discrete path integral and spin foam context. For earlier work on matter coupling (including gauge fields) in the GFT, discrete path integral and spin foam (and loop quantum gravity) frameworks, see \cite{Ambjorn:1992us,Hamber:1993gn,Ambjorn:1993ta,MoralesTecotl:1995jh,Thiemann:1997rt,Oriti:2002bn, Freidel:2004vi, Freidel:2005cg, Oriti:2006jk, Fairbairn:2006dn, Speziale:2007mt, Dowdall:2010ej, Dowdall:2009ds, Bianchi:2010bn, PhysRevD.93.024042} and references cited therein\footnote{Another possibility, that has been only marginally explored in the GFT literature \cite{Fairbairn:2007sv,Girelli:2009yz,Oriti:2009pb}, is that matter degrees of freedom could be emergent from, rather than coupled to, the quantum geometric ones.}. We consider minimally coupled scalar fields, with standard propagators and generic interactions first. Then we consider variations of the corresponding GFT model, leading to generalized scalar field actions. 

\

\noindent Our strategy is the following. We introduce the appropriate additional variables describing scalar matter in the GFT fields and states, thus extending the standard quantum geometry states and spin networks. Then we seek to determine the appropriate kinetic and interaction kernels for the GFT action; our criterion for determining them is that the corresponding Feynman amplitudes should take the form of a simplicial path integral for gravity coupled to a relativistic scalar field. To do so, we will mainly work in the non-commutative flux/metric formulation of GFT and spin foam models \cite{Baratin:2010wi,Baratin:2011hp}, in which the amplitudes take the explicit form of a non-commutative simplicial path integral; this allows us to keep the discrete geometry of the model under control, but at the same time will require special care to handle the complications coming from the non-commutativity of the discrete flux/metric variables. Obviously, the discrete gravity path integral is not uniquely defined, nor is the discretization procedure of the scalar matter (nor is the quantization map to be used for quantizing discrete flux variables \cite{Guedes:2013vi}, which in turn is reflected in various elements of the discretization procedure, as we will discuss). What we will care about is only that our choices and manipulations are compatible with the naive classical and continuum limits of the discrete path integral, and with the general properties of classical simplicial geometries, favoring simplicity of the results over exactness of the discrete model whenever two alternatives are equally valid in terms of these limits.
Having obtained a satisfactory GFT model (and thus a corresponding spin foam model) according to these criteria, we will focus on possible approximations of the same model at the level of the GFT action, determining what these approximate models amount to in terms of the corresponding simplicial gravity-scalar field path integral. The main reason to consider these approximations is that they may turn out to be relevant for the extraction of effective continuum physics from the coupled GFT model, and in fact the simplest of the approximations we will consider turns out to correspond to the way a free massless scalar field has been introduced at the GFT level in \cite{Oriti:2016qtz}. Our central interest in the continuum limit of the coupled model is also the reason why we focus on the GFT level of the description, in particular the classical action, rather than expounding the details of the spin foam expansion for the same model. While detailing only the simplicial path integral form of the Feynman amplitudes of the GFT model, we will give anyway enough information to allow the interested reader to obtain the precise form of the spin foam amplitudes for our model, if needed.

 \

\noindent The paper is organized as follows. In section \ref{sec:GFT}, we introduce the basics of the GFT framework that will be used in the following discussion, explaining in particular how the GFT Feynman amplitudes take the form of simplicial path integrals and spin foam models. In section \ref{sec:SFD}, we discuss the elements of the scalar field discretization: discretization of the scalar matter degrees of freedom, discrete geometric quantities needed for the coupling with gravity, scalar field action. In section \ref{sec:cp1} we present the GFT model for gravity minimally coupled to a single relativistic scalar field, and illustrate both the GFT action and the corresponding discrete path integral. In section \ref{sec:cp2} we consider the models obtained by various approximations of the newly derived one, and the discrete scalar field path integrals they correspond to. In section \ref{sec:ga}, we discuss the straightforward generalization to the multiple (complex) minimally coupled scalar fields; we also briefly discuss the use of the newly derived coupled GFT model to extract effective cosmological dynamics, and compare it briefly with the typical dynamics of gravity coupled to a scalar field as given in quantum cosmology. Section \ref{sec:con} contains our conclusions. We include appendices with further technical details on our construction.

\section{Group field theory and discrete gravity path integrals}\label{sec:GFT}

Most GFT models for Riemannian quantum gravity, like most spin foam models, are based on modifications of the Ooguri models for $4$ dimensional BF theory. They are motivated by the fact that classical 4-dimensional gravity can be expressed as a constrained BF theory. The action is given as Plebanski-Holst action\footnote{The actual Plebanski-Holst action is given as
$S[\omega,B,\Psi] = \int_{\mathcal{M}}\tr\left[\left(\ast+\gamma^{-1}\right)B\wedge F[\omega]\right]+\Psi\cdot S(B)$. By redefining $\left(\ast+\gamma^{-1}\right)B\rightarrow B$, one can get \Ref{eq:S}.  }
\begin{equation}\label{eq:S}
S[\omega,B,\Psi] = \int_{\mathcal{M}}\tr\left(B\wedge F[\omega]\right)+\Psi\cdot S_\gamma(B)
\end{equation}
where $B$ is $\mathfrak{so}(4)$ valued 2-form, $\omega$ is $\mathfrak{so}(4)$ connection 1-form and $\Psi$ is a Lagrange multiplier for the simplicity constraint $S_\gamma(B)$. The variation on $\Psi$ introduces the constraint $S_\gamma(B)=0$ whose solution is $B=\pm\left(\ast+\gamma^{-1}\right) e \wedge e$, where $e$ is the tetrad 1-form\footnote{Depending on the specific was in which the simplicity constraints are defined, there may be another sector of solutions: $B= \pm\left(1+\ast\gamma^{-1}\right) e \wedge e$; clearly, the action takes the same form if this set of solutions is considered instead.}. $\gamma$ is the Barbero-Immirzi parameter. 
At the discrete path integral level, one would like to reproduce the same mechanism, with a BF amplitude (i.e. the exponential of a discretised BF action) constrained by the discrete counterpart of the simplicity constraints, and a suitably modified measure. Details on the discretisation procedure and its result can be found for example in \cite{Alexandrov:2011ab,Perez:2012wv}.
The discrete path integral (DPI) of the theory is based on a triangulation $\Delta$ of $\mathcal{M}$, and its (topologically) dual complex $\Delta*$. The $B$ fields and curvature $F$ are discretised on the triangle $f_{\Delta}$ of the triangulation $\Delta$ and its dual face $f_{\Delta*}$, respectively
\begin{equation}
B_f = \int_{f_{\Delta}} B, \qquad F_f = \int_{f_{\Delta*}} F(\omega)
\end{equation}
Relying on a non-abelian extension of Stokes' theorem, the integral of the curvature $F$ on the dual face $f_{\Delta*}$ can be written as a line integral of the connection $\omega$ on the boundary of $f_{\Delta*}$, and this in turn can be traded for the holonomy along the boundary of the dual face obtained as the product of group-valued parallel transports associated to each link forming such boundary, which become the true independent variables replacing the continuum connection field:
\begin{equation}
F_f = \int_{f_{\Delta*}} F = \oint_{\p f_{\Delta*}}\!\!\omega \approx H_f \equiv \prod_{\ell\in\p f_{\Delta*}} g_\ell = \ex^{F_f}(g_\ell)
\end{equation}
where $\ell$ are the links of the dual graph $\Delta*$ each of which connect two neighbouring tetrahedra that sharing triangle $f$, and the approximation holds in a continuum limit of the triangulation $\Delta$, i.e. under the successive refinement of the same. In other words, the discrete curvature $F_f$ corresponds to the Lie algebra element obtained from the holonomy $H_f$ approximated close to the identity.
Then the DPI can be rewritten as
\begin{equation}\label{BFDPI}
Z^{\Delta}_{\rm G}=\int \!\prod_{f\in \Delta}\!\mathcal{D} B_f  \prod_{\ell\in \Delta*}\dd g_{\ell} ~\delta(S_\gamma(B_f)) ~\ex^{\frac{\ii}{2\hbar \kappa}\sum_{f\in \Delta}\tr(B_f  H_f)},
\end{equation}
where $\hbar$ is the Planck constant, $\kappa$ is the bare Newton constant, 
$\dd g$ indicates the Haar measure on $SO(4)$, and the simplicity constraints are imposed by a suitable discretization of the formal $\delta(S_\gamma(B_f))$, one specific form of which we will give in the following, alongside a specific choice for the integration measure over the fluxes $B_f$. Notice that one could have also used the Lie algebra element $F_f$ in the discrete action, instead of the holonomy element $H_f$; the two choices lead obviously to the same result in the naive continuum limit. We will come back to these two possibilities in the following.

The group field theory formalism \cite{Oriti:2011jm,Krajewski:2012aw,Oriti:2014uga,Oriti2015} allows to derive the same discrete gravity path integral as the Feynman amplitude of a combinatorially non-local action on the group manifold $SO(4)^4$ or on the dual Lie algebra $\mathfrak{so}(4)^4$. The details of one such construction for the Plebanski-Holst theory can be found in \cite{Baratin:2011hp}, but the correspondence between a given discrete path integral and a given GFT action is generic. The formulation in terms of group elements and the one in terms of Lie algebra elements are related by non-commutative Fourier transform \cite{Guedes:2013vi,Baratin:2010wi}. We focus on the formulation in terms of Lie algebra elements because they correspond directly to the flux variables $B_f$ and encode more transparently the discrete geometry underlying the theory. At the same time, their non-commutative nature introduces some formal complications, in particular the need to use $\star$-products between functions of Lie algebra elements, encoding the quantization map chosen to quantize them. Details on these non-commutative tools can be found in \cite{Baratin:2011hp, Guedes:2013vi,Baratin:2010wi}.
The general point is to understand $\Delta*$ (and $\Delta$) as the Feynman diagram of a GFT model with action
\begin{eqnarray}
S_{\rm G}^{\rm GFT}&=&\int [dx^4][dx'^4]\bar{\varphi}(\vec{x})\star P_{\rm G}^{-1}(\vec{x},\vec{x}')\star\varphi(\vec{x}')+\nonumber\\
&&\int\left(\prod_{n=1}^{5}[dx_n^4]\right) V_{\rm G}(\vec{x}_1,\ldots,\vec{x}_5)\star \left(\prod_{n=1}^{5}\varphi(\vec{x}_n)\right)+{\rm c.c.},
\end{eqnarray}
where $\varphi$ is our GFT field living on the $4$ copies of the Lie algebra, integrated over using the Lebesgue measure, and $P_{\rm G}$ and $V_{\rm G}$ are the propagator and vertex kernels\footnote{We have used real fields for simplicity, but the formalism can be immediately extended to complex fields, maintaining the same Feynman amplitudes whenever one uses real interaction kernels ${V}_{\rm G}=\bar{V}_{\rm G}$ and propagator.}. The kinetic and interaction kernels are convoluted with GFT fields and multiplied using $\star$-multiplications. The discrete path integral is the associated Feynman amplitude, constructed by convolution of propagators and vertex kernels. Obviously, the kinetic kernel $P_{\rm G}^{-1}$ can be more or less easy to identify, depending on our choice of propagator $P_{\rm G}$ so the expression above remains quite formal at this stage. Notice also that, if one is only interested in obtaining a certain expression for the Feynman amplitudes (i.e. for the simplicial gravity path integral), the exact functional form of propagator and vertex kernel can be modified with some liberty; in particular, this affects the way simplicity constraints can be implemented in the GFT model. We base our discussion here on the specific construction given in \cite{Baratin:2011hp}, to which we refer for details.

In the end, the discrete path integral $Z^{\Delta}_{\rm G}$, with Lie algebra variables $x$ identified with the discrete fluxes $B_f$, associated with the triangles in the triangulation $\Delta$, and group elements $g_\ell$ being the parallel transport of the gravity connection associated with the dual links in $\Delta*$, becomes the GFT Feynman amplitude obtained by convolution of $P_{\rm G}$
and $V_{\rm G}$, schematically
\begin{eqnarray}\label{PureGravFeynExpan}
Z^{\Delta}_{\rm G}&=&\int_{\Delta} V_{\rm G}(\ldots,\vec{x}^N_i,\ldots,\vec{x}^N_j,\dots)\star P_{\rm G}(\vec{x}^N_i,\vec{x}^{n_1}_{i_1})\star P_{\rm G}(\vec{x}^N_j,\vec{x}^{n_2}_{j_1})\star \ldots \star \nonumber\\
&& V_{\rm G}(\ldots,\vec{x}^{n_2}_{j_1},\ldots)\star \ldots \star  V_{\rm G}(\ldots,\vec{x}^{n_1}_{i_1},\ldots)\star \ldots \star V_{\rm G}(\vec{x}^1_1,\ldots,\vec{x}^1_5)\nonumber\\
&\equiv&\int_{\Delta}\prod_{\vec{\star}}P_{\rm G}V_{\rm G}
\end{eqnarray}
where $i,i_1,j,j_1 \in \{1,\ldots,5\}$, $1<n_1<n_2<N$ and $N$ denotes the total number of the vertex functions in $Z^{\Delta}_{\rm G}$; besides $\vec{x}\equiv(x_1,\ldots,x_4)$ where $x$ is the element of the Lie algebra $\mathfrak{g}$ in the fundamental representation. So since $P_{\rm G}$ and $V_{\rm G}$ are functions of the (non)commutative Lie algebra elements, they are connected by the $\star$-products which, by definition, are performed between each pair of shared arguments of $P_{\rm G}$ and $V_{\rm G}$ as indicated by Eq. (\ref{PureGravFeynExpan}). The precise definition of the $\star$-product for various choices of quantization maps can be found in the literature \cite{Guedes:2013vi,Baratin:2010wi}, but does not concern us here. We use from now on $\int_{\Delta}$ to indicate the total convolution with all the arguments of the propagators and vertex functions associated with $\Delta$, as in a usual Feynman amplitude for a closed Feynman diagram, but omitting the details of the integration measure for the ease of reading. \\

For the GFT/spin foam model in \cite{Baratin:2011hp}, the form of $P_{\rm G}$ and $V_{\rm G}$ are specified as follows. For $P_{\rm G}$, we have
\begin{eqnarray}
P_{\rm G}(\vec{x},\vec{x}')=\int dg \prod_{i=1}^{4}\left(e_g\star\delta^{\star}_{-x'_i}\right)(x_{(4-i)}),
\end{eqnarray}
where the 4-tuple of Lie algebra elements $\vec{x}\equiv(x_1,\ldots,x_4)$ is assigned to a tetrahedron of the simplicial complex $\Delta$, with the indices denoting the four triangles of the tetrahedron whose (intrinsic) geometry is encoded in the Lie algebra elements. The Lie group element $g$ is the parallel transport along the link of the cellular complex $\Delta*$ that is dual to this tetrahedron, and $e_g(x)$ is the plane wave of the Lie group (or algebra), whose definition depends on the specific quantization map one uses for the classical fluxes associated to the triangles (which become the variables $x$). Two common choices are the \textit{Freidel}-\textit{Livine}-\textit{Majid} case or the \textit{Duflo} case \cite{Oriti:2014qoa,Oriti:2014aka}. Finally, $\delta_{-x'_i}^\star(x_i)$ is the \textit{Dirac} $\delta$-function for the given $\star$-product:
\begin{eqnarray}
\int dx \left({\rm f}\star\delta_{-x'_i}^\star\right)(x)={\rm f}({-x'_i}),
\end{eqnarray}
identifying $x_i$ with $-x'_i$ (where the minus sign reflects the opposite orientations of the triangle that is assigned the Lie algebra elements $x_i$ and $x_i'$ respectively, when seen from the two 4-simplices dual to the vertices of $\Delta*$, which are also the interaction vertices of the GFT Feynman diagram) sharing the tetrahedron to which the propagator is associated.
Due to the integral over $g$ and the multiplication by plane wave, $P_{\rm G}$ imposes also the \lq closure constraint\rq for the tetrahedra it is associated with, i.e. it forces the four triangles to which the $x_i$ are associated to close to form its boundary. This is part of the geometricity constraints that allow a geometric interpretation for the variables $x_i$ and for the simplicial complex $\Delta$. They also introduce the gauge connection encoded in the same variables $g$ associated to each dual link. They are not enough, however.

For $V_{\rm G}$, we have
\begin{eqnarray}
V_{\rm G}(\vec{x}_1,\ldots,\vec{x}_5)=\left(\prod_{j=1}^{5}S(\vec{x}_j)\right)\star\prod_{i\neq j}\delta^{\star}_{-x_{ij}}(x_{ji}),
\end{eqnarray}
where the indices $i,j = 1, ..., 5$ denote the five tetrahedra of a $4$-simplex, with which $V_{\rm G}$ is associated.

As anticipated, and in parallel with the continuum Plebanski-Holst action, the $x$'s, interpreted as the discrete counterpart of the $B$ fields of the continuum theory, have too many degrees of freedom to be geometrical, even after the closure condition is imposed on them. We need to impose the discrete counterpart of the simplicity constraints of the continuum theory. This is the role of the functions $S(\vec{x})$ in the vertex kernels $V_{\rm G}$:
\begin{eqnarray}
S(\vec{x})&=&\prod_{i=1}^4 S^\beta_n(x_i) \nonumber\\
&=&\prod_{i=1}^4 \delta^\star_{-n x_i^- n^{-1}}(\beta x_i^+),
\end{eqnarray}
which indeed impose, via non-commutative delta functions under the integration, the simplicity constraint $\beta x^+=-nx^-n^{-1}$, where $\beta=\frac{\gamma-1}{\gamma+1}$ and $n\in S^3\simeq {\rm SU}(2)$ is a free chosen vector representing the normal to the tetrahedra to which each $S$ function is associated \cite{Baratin:2011hp}. This normal vector enters crucially the definition of the GFT model, and should actually be incorporated as an extra argument in the GFT field as well, and coupled (albeit in a very simple way) in the GFT action (see \cite{Baratin:2011hp} for details). The same variables, however, disappear from the discrete path integral, as they can be reabsorbed in the discrete connection variables, and we omit them as well from the GFT fields and action, for simplicity of notation.

This is just the (linear) simplicity constraint for $4$-dimensional Riemannian geometry, using the selfdual/anti-selfdual decomposition of $\mathfrak{so}(4)$ Lie algebra elements $x = ({x}^+,{x}^-)$ as two ${\mathfrak s}{\mathfrak u}(2)$ elements ${x}^+$ and ${x}^-$. The Lorentzian case, based on $SL(2,\mathbb{C})$, is also under control, but we will stick to the simple case of Riemannian gravity in the following.

The product of non-commutative $\delta$-functions of $\vec{x}_j$ enforces the requirement that the five tetrahedra involved in $V_{\rm G}$ are glued pairwise (through the identifications of $\vec{x}_j$ associated to their common triangles) to form a $4$-simplex, so that the dual of the GFT Feynman diagrams correspond to $4$-dimensional simplicial complexes.

\begin{figure}
\includegraphics{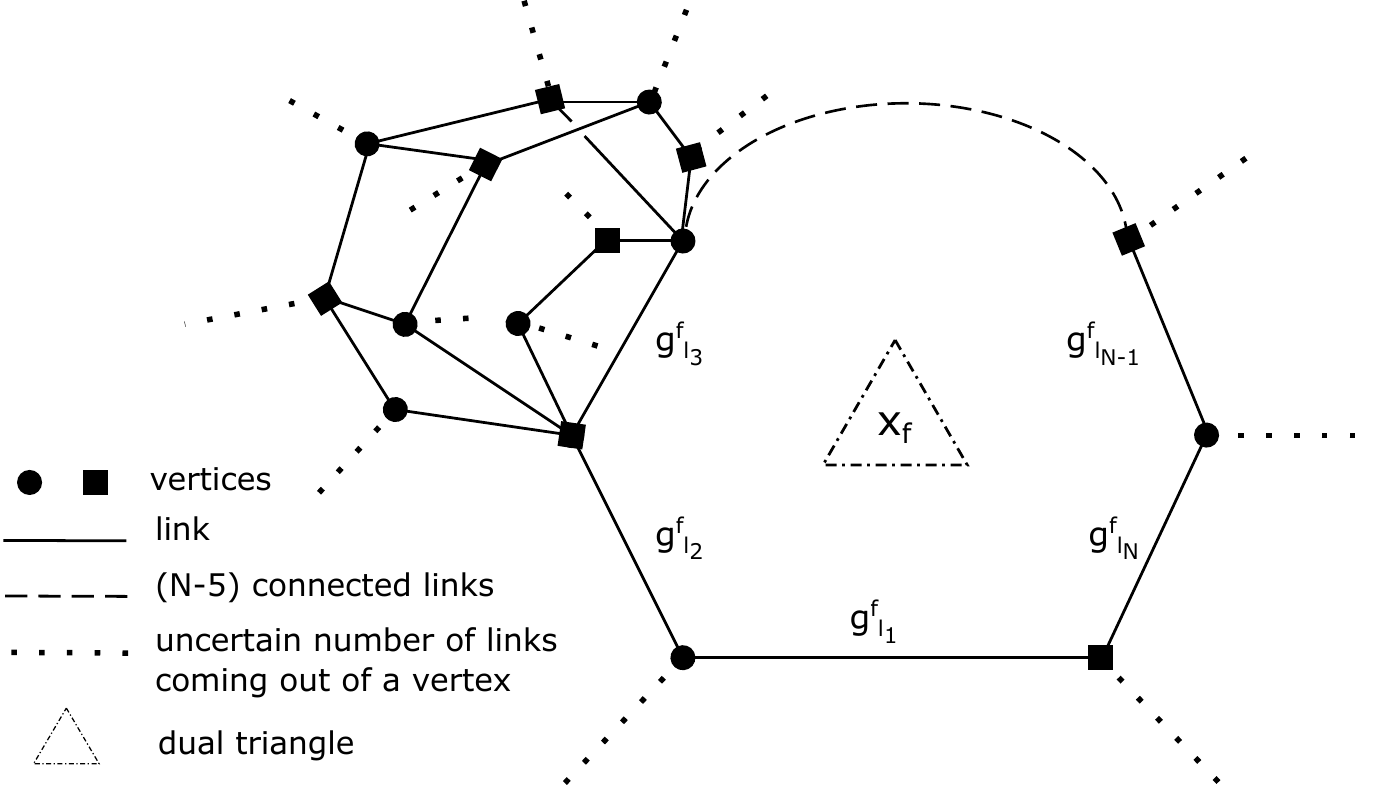}
\centering
\caption{The figure shows a glance of the complex $\Delta*$: every vertex is shared by $5$ links and the two different vertices correspond to $V_{\rm G} = \bar{V}_{\rm G}$. Faces are formed by closed loops of links where every link is shared by $4$ faces. The $N$ links enclose the face $f$ dual to a triangle of the simplicial complex $\Delta$. Every link is associated with $g_l$, and the dual triangle is associated with $x_f$.}
\label{FaceFigure}
\end{figure}

Due to the (non)commutative $\star$-products of $P_{\rm G}$ and $V_{\rm G}$ in Eq. (\ref{PureGravFeynExpan}), it is crucial, in the actual construction of the Feynman amplitudes, to further specify their correct ordering. This ordering relates the different Lie algebra variables associated {\it to the same triangle} in the propagators and vertex kernels associated to the tetrahedra and 4-simplices, respectively, sharing the same triangle, for a given complex $\Delta*$. 
An example of how to define the ordering for a given lattice is shown in the appendix \ref{app:order}. It is crucial to point out that this ordering is just the one produced by the Feynman expansion of the corresponding GFT. \\
\indent By explicit construction, for any choice of Feynman diagram, the reader can convince herself or himself that $Z^{\Delta}_{\rm G}$ can be factorized into the integral of the face amplitude $A_f$:
\begin{eqnarray}
Z^{\Delta}_{\rm G}=\int\left(\prod_{l\in \Delta*} dg_l\right)\left(\prod_{f\in \Delta*}A_f(\vec{g}_f)\right),
\end{eqnarray}
where $\vec{g}_f= \left( g^f_{l_1}, g^f_{l_2}, \ldots, g^f_{l_N}\right)$ denotes the group elements associated  to the face $f$ whose boundary is constituted by $N$ links $l$ and
\begin{eqnarray}\label{FaceAmplit}
A_f(\vec{g}_f)&=&\int dx_f \left(\bigstar_{i=1}^N S^{\beta\star 2}_{g_{1i}\triangleright n} \right)\star e_{H_f}(x_f)\\
\bigstar_{i=1}^N S^{\beta\star 2}_{g_{1i}\triangleright n}&=&S^{\beta\star 2}_{g_{11}\triangleright n} \star S^{\beta\star 2}_{g_{12}\triangleright n}\star \ldots \star S^{\beta\star 2}_{g_{1N}\triangleright n}\\
g_{1i}\triangleright n&:=&g_{1i}^+ n \left(g_{1i}^-\right)^{-1}
\end{eqnarray}
where $x_f\equiv x_1$, $H_f\equiv g^f_{l_1}g^f_{l_2}\ldots g^f_{l_N}$, $g_{1i}=g^f_1 g^f_2\ldots g^f_i$ (with $g_{1i}^+/g_{1i}^-$ corresponding to the selfdual/antiselfdual decomposition) and $S^{\beta\star 2}_{g_{1i}\triangleright n}\equiv \left(S^\beta_{g_{1i}\triangleright n}\star S^\beta_{g_{1i}\triangleright n}\right)(x_f)$ corresponding to the superposition of two simplicity constraints coming from both tetrahedra that share a triangle; see Fig. \ref{FaceFigure}.
We obtain an amplitude which depends on $H_f$, i.e. the holonomy around the face $f$, and $x_f$ (satisfying the simplicity constraints), i.e. the constrainted bivector $B_f$ from Eq. (\ref{BFDPI}). The amplitude can be recognised more explicitly as a discrete gravity path integral (in terms of non-commutative metric variables) if we make explicit the form of the non-commutative plane waves. If we adopt the \textit{Freidel}-\textit{Livine}-\textit{Majid} definition of the plane wave, we have $e_{H_f}^{\rm FLM}(x_f)=e^{\frac{i}{2\hbar \kappa}{\rm Tr}(x_f  H_f)}$; if we adopt the \textit{Duflo} definition of the plane wave, we have $e_{H_f}^{\rm D}(x_f)=e^{\frac{i}{2\hbar \kappa}{\rm Tr}(x_f  F_f)}$. Also, $\dd x_f$ plays the role of $\mathcal{D}B_f$ as well as $\dd g_\ell$ plays the role as $\mathcal{D}\omega$ given the fact they are Haar measures of a compact group. 

So at last we have
\begin{eqnarray}
Z^{\Delta}_{\rm G}=\int\left(\prod_{f\in \Delta*}\dd x_f\right)\left(\prod_{l\in \Delta*} \dd g_l\right)\left(\prod_{f\in \Delta*}\bigstar_{i=1}^N S^{\beta\star 2}_{g_{1i}\triangleright n}\right)e^{\frac{i}{2\hbar \kappa}\sum_{f\in \Delta}{\rm Tr}(x_f  (H_f/F_f))},
\end{eqnarray}
from which we can see how the GFT Feynman amplitude reproduces the DPI of the constrained BF theory for gravity, with $\prod_{f\in \Delta*}\bigstar_{i=1}^N S^{\beta\star 2}_{g_{1i}\triangleright n}$ encoding the simplicity constraints $\delta(S_\gamma(B))$ in Eq. (\ref{BFDPI}) (see again \cite{Baratin:2011hp} for details).

\section{Scalar field discretization}\label{sec:SFD}

In order to couple the scalar field to gravity at the GFT level, we will focus on the discretization of a relativistic real scalar field on a simplicial manifold, and then look for a GFT model that would reproduce it as its Feynman amplitude, thus generalising the pure gravity case we have described in the previous section. We pay special attention to those features that are crucial for the correct (naive) continuum limit.
Let us first write down the continuum action of a real relativistic scalar field on a spacetime $\mathcal{M}$ with topology and geometry captured by a metric $g$:
\begin{eqnarray}
S_{\phi}=\int_{\mathcal{M}} [\dd x^4]\sqrt{|{\rm g}|}\left( \partial_\mu \phi \partial^\mu \phi+{\mathbb V}(\phi)\right)
\end{eqnarray}
where $|{\rm g}|$ is the absolute value of the determinant of the spacetime metric $g$; 
${\mathbb V}(\phi)$ is a generic potential term which characterizes the self-interaction of the scalar field. The continuum path integral is then (formally) defined by the exponential of this action, integrated with a functional measure that is intuitively (but only formally) given by the Lebesgue measure over the values of the scalar field at each point in spacetime.
Now we move to the discretization of this scalar field theory.

The first crucial choice is whether we discretize the scalar field degrees of freedom on the simplicial complex $\Delta$ or on the dual cellular complex $\Delta*$. This is going to affect considerably both the structure of the simplicial path integral and the corresponding GFT action. For reasons that will become obvious in the following, we choose to discretize the scalar field on the dual complex $\Delta*$ and to work with the discrete path integral given as:
\begin{eqnarray}\label{z_s}
Z^{\Delta}_{\phi}&=&\int\!\! \prod_{v\in \Delta*} \!\!\dd\phi_{v}~~ \ex^{\frac{\ii}{\hbar}\left(\sum_{l\in \Delta*}\tilde{V}_{l}\left(\frac{\delta_l \phi}{L_l}\right)^2+\sum_{v\in\Delta*}V_{v}{\mathbb V}(\phi_v)\right)}\qquad .
\end{eqnarray}

Here, $\phi$ (having dimension of inverse length, with $\hbar = 1$) is discretized on the vertices $v$ of $\Delta*$ as $\phi_v$, while its derivatives are discretized as finite differences
$\delta_l \phi\equiv\phi_{{v_l}'}-\phi_{v_l}$ between the values at two vertices ${v_l}$ and ${v_l}'$ and thus associated to the link $l$ in $\Delta*$ connecting them. As $L_l$ denotes the length of this link, $\left(\frac{\delta_l \phi}{L_l}\right)^2$ is thus the discretization of $\partial_\mu \phi \partial^\mu \phi$, while the scalar field potential ${\mathbb V}(\phi_v)$, being a simple polynomial of the field, is also associated to the vertex $v$. Important for us is of course the coupling with the geometry of the simplicial complex (and its dual) encoded in the discrete metric variables. In particular, we need the discretization of the continuum volume element. This has to be done differently in the kinetic and interaction terms. We have defined $\tilde{V}_{l}$ to be the volume of the $4$-dimensional convex hull whose vertices are ${v_l}$, ${v_l}'$ and the vertices of the tetrahedron dual to the link $l$; $V_{v}$ is instead the volume of the $4$-simplex dual to the vertex $v$. These two volumes, adapted to the functions of scalar field variables that multiply them,  play the role of discrete $4$-dimensional volume element in the kinetic term and the potential term respectively. 


Because the quantities defining the coupling of the discretized scalar field with the simplicial geometry depend both on the triangulation $\Delta$ and on its dual complex $\Delta*$, their precise evaluation requires to specify the embedding of the second into the first, so that the geometry of the second can be deduced by the piecewise-flat geometry of the first. An extended discussion of the possible definitions and their consequences can be found in \cite{Calcagni2013}. Different embedding choices result in different expressions for the geometric observables entering the discretized scalar field action and thus, in the end, in different GFT models. However, they are all expected to be equivalent in a naive continuum limit of the simplicial path integral, as they approximate each other in a refinement limit and for regular enough triangulations; this is also the only condition we really require on our construction, and in the following, when facing ambiguities in the model building, we will repeatedly look for the construction which gives the simplest result compatible with this main requirement.

Here we choose each vertex $v$ of the dual complex $\Delta*$ to be the center of the sphere inscribed in its dual 4-simplex.
The geometrical quantities $\tilde{V}_{l}$, $L_l$ and $V_{v}$ can all be written as functions of the Lie algebra elements $x$ (understood as constrained bivectors) encoding the simplicial geometry of the pure gravity GFT model, and in particular the dual volumes can be expressed as functions of geometric volumes of elements of the triangulation, as follows:
\begin{eqnarray}
\tilde{V}_{l}&=&\frac{|V_{(3)l}|}{\bar{V}_{(3)v_l}}V_{v_l}+\frac{|V_{(3)l}|}{\bar{V}_{(3){v_l}'}}V_{{v_l}'}\\
L_l&\geq&\tilde{L}_l=\frac{4|V_{v_l}|}{\bar{V}_{(3)v_l}}+\frac{4|V_{{v_l}'}|}{\bar{V}_{(3){v_l}'}}\\
V_{v}&=&\frac{1}{24}\epsilon_{QIJK}\tilde{x}'^{QI}_v\tilde{x}^{JK}_v
\end{eqnarray}
with
\begin{eqnarray}
|V_{(3)}|&=&\frac{1}{6\sqrt{2}}|X^{\frac{1}{2}}|\\
X&=&\sum_{i,j,k,q}\frac{1}{2}\epsilon^{ijkq}\epsilon_{QIJK}\tilde{x}_{i}^{QI}\tilde{x}_{j}^{JM}\tilde{x}_{{k}M}^{\,\,\,\,\,\,\,\,\,\,\,\,K} \qquad .
\end{eqnarray}
Here $\tilde{x}=\frac{1}{\gamma^2 - 1}(\gamma^2 x- \gamma (*x))$ and $(*x)^{QI}=\frac{1}{4}\epsilon_{QIJK}x^{JK}$, where the capital Latin letter indices are ${x}$'s bivector indices (or matrix indices in the fundamental representation).
The lower case indices denote four triangles of a tetrahedron which are summed from $1$ to $4$. Then $|V_{(3)l}|$ is the modulus of the $3$-volume of the tetrahedron dual to the link $l$; and $\bar{V}_{(3)v}$ is the sum of the moduli of the volumes of the boundary tetrahedra of the $4$-simplex dual to the vertex $v$. For each $V_{v}$, $\tilde{x}_v$ and $\tilde{x}'_v$ are associated with any two triangles on the boundary of the $4$-simplex dual to vertex $v$ that do not share a common edge. We do not give the exact expression of $L_l$, which is quite involved and not particularly illuminating, but only its lower limit $\tilde{L}_l$, which is the sum of the two radiuses of the inscribed spheres of two neighbouring $4$-simplex dual to ${v_l}$ and ${v_l}'$, respectively. More details on these quantities can be found in the appendix \ref{app:VLV}. 

\noindent The above quantities complete the definition of the discrete action (and path integral) for gravity coupled to a scalar field. Quite clearly, they are rather involved, and so would be the corresponding GFT action. Therefore, we write down a simplified version of $\tilde{V}_{l}$, $L_l$ and $V_{v}$, in addition to their exact expressions and, in the following, we will mainly use these simplified expressions. The reason is again that we mainly aim at capturing the correct scalar field coupling at the effective continuum limit, and thus at the discrete level (which is never uniquely specified by the continuum theory, in any case) the main requirement is that our model has the correct naive continuum limit (i.e. it approximate the continuum action for triangulations that are sufficiently regular, with small edge lengths and weak curvature \cite{Hamber:1993gn}) while being as simple (and manageable) as possible. In determining the most convenient simplification of the exact expressions, we are also keep in mind our goal of obtaining a manageable GFT model. This will be much simpler if the geometric couplings used can refer exclusively to the geometric variables associated to a single tetrahedron, the only ones that appear in the GFT kinetic term, and to a single 4-simplex, the only ones appearing in a GFT interaction term.


\noindent With these criteria in mind, we are going to use the following simplified version of $\tilde{V}_{l}$, $L_l$ and $V_{v}$:
\begin{eqnarray}\label{VLV}
\tilde{V}&=&{\rm Sgn}\left(X\right)\frac{1}{48\sqrt{5}}|X^{\frac{2}{3}}|\label{tildeV}\\
L&=&\frac{1}{\sqrt{10}}|X^{\frac{1}{6}}|\label{L}\\
V&=&{\rm Sgn}\left(X\right)\frac{\sqrt{5}}{96}|X^{\frac{2}{3}}|\label{V} \qquad .
\end{eqnarray}

\noindent The labels $l$ and $v$ for $\tilde{V}$, $L$ and $V$ are neglected for ease of notation. 
Here, $\tilde{V}$, $L$ are all functions of the fluxes $x$ from a single tetrahedron dual to the link $l$. We have also reported a (drastically) simplified expression for $V$, in terms of the data from one of the tetrahedra on the boundary of the $4$-simplex it refers to, although for $V$ we do not really need such simplification and could work with the exact expression. Once more, we would expect a discrete path integral (and a GFT model) with these simplified expressions used instead of the exact ones to produce the same effective continuum physics, at least in a macroscopic, semiclassical approximation.

\noindent The ${\rm Sgn}\left(X\right)$ in $\tilde{V}$ and $V$ make sure they have consistent orientations as discrete volume elements. For $L$, the orientation is not important since it will always appear under the square (coming from the discretization of $\partial_\mu \phi \partial^\mu \phi$). The details of the procedure leading to these simplifications can be found in the appendix \ref{app:VLV}.\\
\noindent In addition, since $\tilde{V}$, $L$ and $V$ (simplified or not) are only a function of the discrete geometry of the triangulation $\Delta$, they actually depend only on the edge lengths of the simplicial complex (since its discrete geometry can be defined solely in terms of the edge lengths). In our $4$-dimensional context, these geometrical functions can equally be written purely as a function of the triangle areas, which can also define the discrete geometry of the $4$-dimensional simplicial complex (uniquely, when subject to appropriate additional constraints \cite{Makela:2000ej, Calcagni:2012cv}). So $\tilde{V}$, $L$ and $V$ only depend on $|\tilde{x}|\propto |{x}|$.

\noindent Finally, we point out that one could be interested in more general continuum actions for the relativistic scalar field, with kinetic terms of the general form ${\mathbb K}(\partial_\mu \phi \partial^\mu \phi)$, where ${\mathbb K}$ is a general real function. Our discretization procedure allows to deal easily with these generalised case as well: we just need to replace $\left(\frac{\delta_l \phi}{L}\right)^2$ with ${\mathbb K}\left((\frac{\delta_l \phi}{L})^2\right)$ in Eq. (\ref{z_s}). 

\section{GFT with minimal coupled real scalar field}\label{sec:cp1}
In this section we give the GFT model which has the discrete path integral for gravity minimally coupled to a real scalar field, as introduced in the previous sections, as its Feynman amplitude. Let us first write down such discrete path integral:
\begin{eqnarray}\label{PartiFuncSplit}
Z^{\Delta}_{\phi\rm G} &=& \int \prod_{f\in \Delta}\mathcal{D}B_f \prod_{f\in \Delta*}\dd g_{\ell} \prod_{v\in \Delta*} \dd\phi_{v} ~\delta(S_{\gamma}(B_f))~\ex^{\frac{i}{\hbar}\left( S_\phi^\Delta+S^\Delta_G\right)}
\end{eqnarray}
where
\begin{equation}
S_\phi^\Delta \equiv \left(\sum_{l\in \Delta*}\tilde{V}_{l}\left(\frac{\delta_l \phi}{L_l}\right)^2+\sum_{v\in\Delta*}V_{v}{\mathbb V}(\phi_v)\right)
\end{equation}
and
\begin{equation}
S_G^\Delta \equiv \frac{1}{2 \kappa}\sum_{f\in \Delta}\tr(B_f  H_f) \qquad ,
\end{equation}

where one can also use the discrete curvature, rather than its holonomy, in the definition of the discrete gravity action.

As in the pure gravity case, in order to incorporate it within the GFT framework, we need to rewrite $Z^{\Delta}_{\phi\rm G}$ as a convolution of propagators and vertex functions, i.e. terms that can be associated to links and vertices of $\Delta*$. Since we already know how to handle $Z^{\Delta}_{\rm G}$, a possible way to do this is to factorize $\exp(\frac{i}{\hbar} S_\phi^\Delta)$ in terms of propagators $P_{\phi}$ and vertex functions $V_{\phi}$, and then to move them inside the pure gravity part of $Z^{\Delta}_{\phi\rm G}$ along with $P_{\rm G}$ and $V_{\rm G}$ to form the complete propagator $P_{\phi\rm G}$ and vertex function $V_{\phi\rm G}$ for $Z^{\Delta}_{\phi\rm G}$.\\
\indent So we could start by assuming:
\begin{eqnarray}
P_{\phi}&=&e^{\frac{i}{\hbar}\tilde{V}\left(\frac{\delta_l\phi}{L}\right)^2}\\
V_{\phi}&=&e^{\frac{i}{\hbar}V{\mathbb V}(\phi_v)},
\end{eqnarray}
where we use our simplification in Eq. (\ref{VLV}), $P_{\phi}$ only depends on the variables $\tilde{V}$ and $L$ which relate to the tetrahedron dual to the edge $l$. 
So $P_{\phi}$ depends on the variables associated with a tetrahedron of the triangulation or its dual link, just like $P_{\rm G}$. In the same way, $V_{\phi}$ can be associated with a $4$-simplex of the triangulation or its dual vertex, just like $V_{\rm G}$.

However we should recall that $\tilde{V}$, $L$ and $V$ are geometrical functions of Lie algebra element $x$, which are non-commutative variables. If we want to treat them as ordinary functions, $P_{\phi}$ and $V_{\phi}$ should be connected via $\star$-products. This has two consequences. First of all, if we multiply all the $P_{\phi}$ and $V_{\phi}$ associated with $\Delta*$ with the noncommutative $\star$-products (no matter what their order is), the result will not simply be $\exp(\frac{i}{\hbar}S_\phi^\Delta)$. Second, even if $\exp(\frac{i}{\hbar} S_\phi^\Delta)$ could be $\star$-product factorized with $P_{\phi}$ and $V_{\phi}$, we can not move them freely through $P_{\rm G}$ and $V_{\rm G}$ (in the gravity part of $Z^{\Delta}_{\phi\rm G}$) without producing additional terms ($\star$-commutators); these additional terms on the one hand may fail to leave the whole amplitude a simple convolution of terms easily identifiable with GFT propagators and vertices, and, on the other hand, will certainly complicate the expression considerably and bring it away from the simple form we have identified for the discrete path integral.

\indent However, we recall once more that we are only interested in obtaining a GFT Feynman amplitude that is equivalent to the discrete path integral we use as reference in a continuum limit, and up to additional terms in the measure or the discrete action that can be interpreted as quantum corrections (the difference between the discrete path integral using in the classical action directly the discrete curvature and the one using its holonomy is exactly of this type). Given this, the general properties of the $\star$-product come to rescue. On the other hand, we know for the general functions of $x$: ${\rm f}(x)$ and ${\rm f}'(x)$, the $\star$-product will reduce to the usual point-wise product in the semiclassical limit $\hbar \rightarrow 0$:
\begin{eqnarray}\label{StarProdProp1}
({\rm f}\star {\rm f}')(x)={\rm f}(x){\rm f}'(x)(1+{\rm O}(\hbar)),
\end{eqnarray}
where, for any given definition of the plane wave, one can explicitly compute the correction ${\rm O}(\hbar)$ (more details on this point can be found in the appendix \ref{app:correction}). This also means for the $\star$-product commutator:
\begin{eqnarray}\label{StarProdProp2}
[{\rm f},{\rm f}']_{\star}(x)\equiv({\rm f}\star {\rm f}')(x)-({\rm f}'\star {\rm f})(x)={\rm O}(\hbar).
\end{eqnarray}

\noindent Knowing these properties, we can approach our problem from another end, and assume that the whole propagator $P_{\phi\rm G}$ and vertex function $V_{\phi\rm G}$ have the following structure:
\begin{eqnarray}
P_{\phi\rm G}(\vec{x},\vec{x}';\phi,\phi')&=& P_{\phi}(\vec{x};\phi,\phi')\star P_{\rm G}(\vec{x},\vec{x}')\\
V_{\phi\rm G}(\vec{x}_1,\ldots,\vec{x}_5;\phi)&=& V_{\phi}(\vec{x}_1,\ldots,\vec{x}_5;\phi) \star V_{\rm G}(\vec{x}_1,\ldots,\vec{x}_5),
\end{eqnarray}
where we already take $V_{\phi\rm G}$ as a local interaction of $\phi$.\\
\indent Then given the cellular complex $\Delta*$, we can write down the Feynman amplitude $\tilde{Z}^{\Delta}_{\phi\rm G}$ by simply substituting $P_{\phi\rm G}$ and $V_{\phi\rm G}$ for $P_{\rm G}$ and $V_{\rm G}$ in Eq. (\ref{PureGravFeynExpan}). This ensures that, if we switch off the scalar field, we obtain back the pure gravity discrete path integral:
\begin{eqnarray}
\tilde{Z}^{\Delta}_{\phi\rm G}&=&\int_{\Delta*} V_{\phi\rm G}(\ldots,\vec{x}^N_i,\ldots,\vec{x}^N_j,\ldots;\phi_N)\star P_{\phi\rm G}(\vec{x}^N_i,\vec{x}^{n_1}_{i_1};\phi_N,\phi_{n_1})\star  P_{\phi\rm G}(\vec{x}^N_j,\vec{x}^{n_2}_{j_1};\phi_N,\phi_{n_2})\star \ldots \star  \nonumber\\
&&V_{\phi\rm G}(\ldots,\vec{x}^{n_2}_{j_1},\ldots;\phi_{n_2})\star\ldots \star V_{\phi\rm G}(\ldots,\vec{x}^{n_1}_{i_1},\ldots;\phi_{n_1})\star \ldots \star V_{\phi\rm G}(\vec{x}^1_1,\ldots,\vec{x}^1_5;\phi_1)\nonumber\\
&\equiv&\int_{\Delta*}\prod_{\vec{\star}}P_{\phi\rm G}V_{\phi\rm G},
\end{eqnarray}
where in $\int_{\Delta*}$, all the arguments of $P_{\phi\rm G}$ and $V_{\phi\rm G}$, including $\phi$, are convoluted.\\

\noindent In order to show that the Feynman amplitude so defined coincides (up to quantum corrections and discretization-dependent terms) with the coupled discrete path integral, i.e. that $\tilde{Z}^{\Delta}_{\phi\rm G}=Z^{\Delta}_{\phi\rm G}$, the following steps have to be taken:
\begin{enumerate}
\item We separate $P_{\phi}$ and $V_{\phi}$ from $P_{\rm G}$ and $V_{\rm G}$ by moving all $P_{\phi}$ and $V_{\phi}$ to the end of the expression with the order among all $P_{\rm G}$ and $V_{\rm G}$ unchanged. 
\item For all the $\star$-products involving $P_{\phi}$ and $V_{\phi}$ in the resulting expression, we change them to the normal products.
\end{enumerate}
In detail, for step $1$, we use Eq. (\ref{StarProdProp2}):
\begin{eqnarray}
\tilde{Z}^{\Delta}_{\phi\rm G}&=&\int_\Delta(\ldots) \star \left(P_{\phi}\star V_{\rm G}\right)(x)\star (\ldots)\nonumber\\
&=&\int_\Delta(\ldots) \star \left(\left(V_{\rm G}\star P_{\phi}\right)(x)+[P_{\phi},V_{\rm G}]_\star(x)\right)\star (\ldots)\nonumber\\
&=&\int_\Delta\left((\ldots) \star \left(V_{\rm G}\star P_{\phi}\right)(x) \star (\ldots)+ (\ldots) \star [P_{\phi},V_{\rm G}]_\star(x) \star (\ldots)\right) \nonumber \\
&=&\int_\Delta\left((\ldots) \star \left(V_{\rm G}\star P_{\phi}\right)(x) \star (\ldots)+ {\rm O}(\hbar)\right),
\end{eqnarray}
where we show a substep of step $1$ by explicitly moving one $P_{\phi}$ through one $V_{\rm G}$ without touching the other unrelated terms. The $\star$-product commutator is produced as expected but it only contributes an ${\rm O}(\hbar)$ term, and this is also true if we move $P_{\phi}$ or $V_{\phi}$ through $P_{\rm G}$ or $V_{\rm G}$. After completing step $1$, one eventually gets
\begin{eqnarray}
\tilde{Z}^{\Delta}_{\phi\rm G}&=&\int_{\Delta}\left(\left(\prod_{\vec{\star}}P_{\rm G}V_{\rm G}\right)\star\left(\prod_\star P_{\phi}V_{\phi}\right)+{\rm O}(\hbar)\right)\nonumber\\
&=&\int_{\Delta}\left(\prod_{\vec{\star}}P_{\rm G}V_{\rm G}\right)\star\left(\prod_\star P_{\phi}V_{\phi}\right)\left(1+{\rm O}(\hbar)\right),
\end{eqnarray}
where $\prod_{\vec{\star}}P_{\rm G}V_{\rm G}$ is as defined before, and $\prod_\star P_{\phi}V_{\phi}$ denotes the $\star$-products of all $P_{\phi}$ and $V_{\phi}$ in some given order.\\
\indent Then using Eq. (\ref{StarProdProp1}), we carry on step $2$ and we have:
\begin{eqnarray}
\tilde{Z}^{\Delta}_{\phi\rm G}&=&\int_{\Delta}\left(\prod_{\vec{\star}}P_{\rm G}V_{\rm G}\right)\star\left(\prod_\star P_{\phi}V_{\phi}\right)\left(1+{\rm O}(\hbar)\right)\nonumber\\
&=&\int_{\Delta}\left(\prod_{\vec{\star}}P_{\rm G}V_{\rm G}\right)\left(\prod P_{\phi}V_{\phi}\right)\left(1+{\rm O}(\hbar)\right)\nonumber\\
&=&\int \prod_{f\in \Delta}\mathcal{D}B_f \prod_{f\in \Delta*}\dd g_{\ell} \prod_{v\in \Delta*} \dd\phi_{v}~~\ex^{\frac{i}{\hbar}S^{\Delta}_{\rm G}}\ex^{\frac{i}{\hbar}S^{\Delta}_{\phi}}+ {\rm O}(\hbar)\,
=\,Z^{\Delta}_{\phi\rm G} + {\rm O}(\hbar).
\end{eqnarray}

Let us point out that, if we use for the geometric observables defining the coupling of the scalar field their expressions as functions only of the modulus of the fluxes, i.e. the areas of the triangles they are associated to (after the imposition of the simplicity constraints), the above manipulations would be much easier, and the result simpler. This is because the commutation between generic functions of the fluxes and plane waves is very simple: ${\rm f}(x)\star e_g(x)=e_g(x)\star {\rm f}(gxg^{-1})$, where $|gxg^{-1}|=|x|$, producing just a rotation of the same fluxes. Therefore, any function that depends only on the modulus of the flux, which invariant under the conjugate action of the group, would be left invariant.

\

\noindent Having identified the propagator and vertex functions $P_{\phi\rm G}$ and $V_{\phi\rm G}$ that produce the desired discrete path integral upon convolution, up to quantum corrections and with some inevitable discretization ambiguities, we can write the corresponding GFT action for gravity minimally coupled to a scalar field. It has the following form:
\begin{eqnarray}\label{CouplGFTAction}
S_{\phi\rm G}^{\rm GFT}&=&\frac{1}{2}\int [\dd x^4][\dd x'^4]\dd \phi \dd \phi'~~{\psi}(\vec{x};\phi)\star P_{\phi\rm G}^{-1}(\vec{x},\vec{x}';\phi,\phi')\star\psi(\vec{x}';\phi')+\nonumber\\
&&\int \dd\phi\left(\prod_{n=1}^{5}[\dd x_n^4]\right)V_{\phi\rm G}(\vec{x}_1,\ldots,\vec{x}_5;\phi)\star\left(\prod_{n=1}^{5}\psi(\vec{x}_n;\phi)\right) \quad +\quad \text{c.c.} \quad ,
\end{eqnarray}
where we have extended the domain of our GFT field $\psi(\vec{x};\phi)$ to include the scalar field degrees of freedom in addition to the metric variables.

Here we use real GFT fields but again the  construction is immediately generalised to complex fields. Notice also that we need to add to the interaction terms their complex conjugate (in absence of additional restrictions on the GFT fields) because $V_{\phi\rm G}\neq\bar{V}_{\phi\rm G}$. The use of one or the other vertex kernel in the construction of the Feynman amplitudes does not affect the form of the simplicial path integral, since the metric variables are integrated symmetrically over the two ranges corresponding to positive and negative volume elements (i.e. opposite orientation of the simplicial manifold), just like in the pure gravity case. Later we will show the kinetic term is real (provided an extra symmetry condition is imposed on the GFT fields).\\

\

\noindent To give a precise expression for the coupled GFT action, we need to find $P_{\phi\rm G}^{-1}$.

\noindent We look first for functions of the form
\begin{eqnarray}
P_{\phi\rm G}^{-1}(\vec{x},\vec{x}';\phi,\phi')=P_{\rm G}^{-1}(\vec{x},\vec{x}')\star P_{\phi}^{-1}(\vec{x}';\phi,\phi'),
\end{eqnarray}
where the component depending on the scalar field variables depends only on the metric variables associated to one tetrahedron in the triangulation (identified across the two 4-simplices sharing it) and satisfies
\begin{eqnarray}\label{CorrInvern}
\int \dd\phi'P_{\phi}^{-1}(\vec{x}';\phi,\phi')\star P_{\phi}(\vec{x}';\phi',\phi'')=\delta(\phi,\phi'').
\end{eqnarray}
Under this assumption, we have
\begin{eqnarray}
&&\int [\dd x'^4]\dd \phi'P_{\phi\rm G}^{-1}(\vec{x},\vec{x}';\phi,\phi')\star P_{\phi\rm G}(\vec{x}',\vec{x}'';\phi',\phi'')\nonumber\\
&=&\int [\dd x'^4]\dd \phi'P_{\rm G}^{-1}(\vec{x},\vec{x}')\star P_{\phi}^{-1}(\vec{x}';\phi,\phi')\star P_{\phi}(\vec{x}';\phi',\phi'')\star P_{\rm G}(\vec{x}',\vec{x}'')\nonumber\\
&=&\delta(\phi,\phi'')\prod_{n=1}^4\delta^\star_{x_n}(x''_n)\qquad ,
\end{eqnarray}
so our assumption is compatible with the defining property of the inverse function.

It is no easy task to find the explicit expression for $P_{\phi}^{-1}$, due to the $\star$-product involved in the above definition. On the other hand, we can once more recall that we are interested in a GFT action that would reproduce the desired discrete path integral only up to quantum corrections, and that the $\star$-product differs in fact from the ordinary point-wise product only by ${\rm O}(\hbar)$ terms. So, if we replace $P_{\phi}^{-1}$ with $\tilde{P}_{\phi}^{-1}$:
\begin{eqnarray}
\tilde{P}^{-1}_{\phi\rm G}=P_{\rm G}^{-1}\star \tilde{P}_{\phi}^{-1},
\end{eqnarray}
where
\begin{eqnarray}\label{SimInvern}
\int \dd \phi'\tilde{P}_{\phi}^{-1}(\vec{x}';\phi,\phi') P_{\phi}(\vec{x}';\phi',\phi'')=\delta(\phi,\phi''),
\end{eqnarray}
the corresponding propagator $\tilde{P}_{\phi\rm G}$ will also differ from $P_{\phi\rm G}$ only by ${\rm O}(\hbar)$ terms:
\begin{eqnarray}\label{tildeP_P}
\tilde{P}_{\phi\rm G}=P_{\phi\rm G}(1+{\rm O}(\hbar)),
\end{eqnarray}
which means the theory will still have the right semiclassical limit at the discrete level. From now on, we will look for the kinetic term $\tilde{P}^{-1}_{\phi\rm G} \backslash \tilde{P}^{-1}_{\phi}$ instead of $P^{-1}_{\phi\rm G}\backslash P^{-1}_{\phi}$.\\
\noindent The above is independent on any specific choice of quantization map for the fluxes, and thus of any special property of the $\star$-product. However, we notice that, if we adopt the \textit{Duflo} quantization map and the corresponding definition of the plane waves, we have in general that
\begin{eqnarray}
\int \dd x ~{\rm f}(x)\star{\rm f}'(x)=\int \dd x ~{\rm f}(x){\rm f}'(x),
\end{eqnarray}
which implies
\begin{eqnarray}
&&\int [\dd x'^4]\dd \phi'\tilde{P}^{-1}_{\phi\rm G}(\vec{x},\vec{x}';\phi,\phi')\star P_{\phi\rm G}(\vec{x}',\vec{x}'';\phi',\phi'')\nonumber\\
&=&\int [\dd x'^4]\dd \phi'\tilde{P}^{-1}_{\phi\rm G}(\vec{x},\vec{x}';\phi,\phi') P_{\phi\rm G}(\vec{x}',\vec{x}'';\phi',\phi'')\nonumber\\
&=&\delta(\phi,\phi'')\prod_{n=1}^4\delta^\star_{x_n}(x''_n).
\end{eqnarray}
\noindent In other words, $\tilde{P}^{-1}_{\phi\rm G}$ is exactly the inverse of $P_{\phi\rm G}$ in this case. As discussed above already, choosing the \textit{Freidel}-\textit{Livine}-\textit{Majid} or the \textit{Duflo} definition for the plane wave is equivalent to choosing different quantization maps for the same classical theory, and results in discrete path integrals which agree in the classical limit. Since this is the only aspect of the theory on which we have solid control, we have no physical reason to prefer one over the other at this stage.
Therefore, also for this reason, we focus on determining $\tilde{P}^{-1}_{\phi\rm G}$ in the following.\\
Here one can also notice that since $P_{\phi}\rightarrow \bar{P_{\phi}}$ when $\tilde{V}\rightarrow -\tilde{V}$, both $P^{-1}_{\phi}$ and $\tilde{P}^{-1}_{\phi}$ will go to its complex conjugate when $\tilde{V}$ flips sign, just like the pure gravity part $P_{\rm G}\backslash P_{\rm G}^{-1}$ does. So now in order to make the kinetic term real under the integral of $x$, one only needs to impose a symmetry of the group field $\psi$ as $\psi(\vec{x};\phi)=\pm\bar{\psi}(\vec{x}';\phi)$ when $\tilde{V}(\vec{x})=-\tilde{V}(\vec{x}')$.

\noindent It is relatively easy to find $\tilde{P}_{\phi}^{-1}$, since $P_{\phi}$ actually only depends on $|\phi-\phi'|$. The simplest way to do it is to express $P_{\phi}$ to the $p$ space with the Fourier transform:
\begin{eqnarray}
P_{\phi}^p&=&\int \dd (\phi-\phi') e^{ip(\phi-\phi')} e^{\frac{i \tilde{V}}{\hbar }\left(\frac{\phi-\phi'}{L}\right)^2}
= \left({\frac{i\pi \hbar L^2}{\tilde{V}}}\right)^{\frac{1}{2}}e^{-i\frac{\hbar L^2}{4\tilde{V}}p^2};
\end{eqnarray}
and
\begin{eqnarray}
\left({P_{\phi}^p}\right)^{-1}= \left({\frac{i\pi \hbar L^2}{\tilde{V}}}\right)^{-\frac{1}{2}}e^{i\frac{\hbar L^2}{4\tilde{V}}p^2} \qquad .
\end{eqnarray}
From this, we easily get $\tilde{P}_{\phi}^{-1}$ by replacing $p$ with $i\partial_{\phi}$:
\begin{eqnarray}\label{NormalScalAct}
\tilde{P}_{\phi}^{-1}&=&\left({\frac{i\pi \hbar L^2}{\tilde{V}}}\right)^{-\frac{1}{2}}e^{-i\frac{\hbar L^2}{4\tilde{V}}\partial_{\phi}^2}
=\sqrt{\frac{ \tilde{V}}{i\pi \hbar L^2}}(1-i\frac{\hbar L^2}{4\tilde{V}}\partial^2_{{\phi}}-\frac{1}{2}(\frac{\hbar L^2 }{4\tilde{V}})^2\partial^4_{{\phi}}+\ldots) \quad .
\end{eqnarray}
\noindent In the end, by replacing $P^{-1}_{\phi\rm G}$ with $\tilde{P}^{-1}_{\phi\rm G}(\vec{x},\vec{x}';\partial^2_\phi)=P_{\rm G}^{-1}(\vec{x},\vec{x}')\star \tilde{P}_{\phi}^{-1}(\vec{x}';\partial^2_\phi)$ in RHS of the Eq. (\ref{CouplGFTAction}), we can write down our GFT action for gravity and a minimally coupled scalar field as:
\begin{eqnarray}\label{TildeS_sG}
\tilde{S}_{\phi\rm G}^{\rm GFT}&=&\frac{1}{2}\int [\dd x^4][\dd x'^4]d\phi {\psi}(\vec{x};\phi)\star \tilde{P}_{\phi\rm G}^{-1}(\vec{x},\vec{x}';\partial^2_\phi)\star\psi(\vec{x}';\phi)+\nonumber\\
&&\int \dd \phi\left(\prod_{n=1}^{5}[\dd x_n^4]\right)V_{\phi\rm G}(\vec{x}_1,\ldots,\vec{x}_5;\phi)\star\left(\prod_{n=1}^{5}\psi(\vec{x}_n;\phi)\right)\quad +\quad \text{c.c.},
\end{eqnarray}
so that we can rest assured that its Feynman expansion will produce the correct discretized path integral for gravity coupled to a scalar field.

\noindent Let us point out how crucial was the choice of discretizing the scalar field on the vertices of the dual complex $\Delta*$, in order to obtain such simple result. In fact, it was this choice that allowed to have a local coupling of the GFT fields with respect to the discrete scalar field variables at the level of the GFT interactions; this would have been impossible if the same scalar field had been discretized on the vertices of the triangulation $\Delta$.

\section{Other GFT models incorporating a real scalar field coupling}\label{sec:cp2}
In the previous section, we have defined a GFT model reproducing the discrete path integral of gravity minimally coupled with a relativistic real scalar filed. However, the procedure we have adopted, basically a form of  "reverse engineering" from the discrete path integral, is at odds with the understanding of GFT as a more fundamental quantum theory of gravity than the one simply defined by the discrete gravity path integral. It would be more natural to simply focus on the theory defined by the GFT action, identify what the more sensible coupling of matter and geometry is at that level, and only at a second stage investigate what discrete theory it corresponds to. In this section, we look at different GFT models defined by actions involving finite derivatives with respect to the scalar field variables, and then investigate the form of their corresponding Feynman amplitudes. The guiding idea is to modify only the propagator of our GFT model, expecting this to correspond to scalar field dynamics with the same local self-interaction as the one we have already studied, but kinetic terms of a more general, but still relativistic type. These models can be seen as approximations of the GFT model we have constructed in the previous section, with the approximations leading to them being rather natural from a GFT point of view, as we will discuss. However, they can also be seen as independent models, and equally interesting from the point of view of quantum gravity. 

\

\noindent In our framework, the scalar field $\phi$ is discretized on the vertices of the $2$-complex, and propagates along the links connecting these vertices. The part of GFT vertex function ${\mathcal V}_{\phi}$ dependent on the scalar field variables encodes the local scalar field potential ${\mathbb V}$ discretized on the same vertices.
On the other hand, ${P}_{\phi}$ encodes the kinetic term of the scalar field in the discrete path integral. A general free dynamics for a relativistic real scalar field would be encoded  in a real function of $\partial_\mu\phi\partial^\mu\phi$: ${\mathbb K}(\partial_\mu\phi\partial^\mu\phi)$.  We look for GFT propagators that would encode this type of generalised scalar field dynamics, corresponding to a (scalar field part of) GFT kinetic kernel $\tilde{\mathcal P}_{\phi}^{-1}(\partial_{\phi}^2)$, where we have omitted the $x$ dependence.
The $\tilde{P}_{\phi}^{-1}$ we get from Eq. (\ref{NormalScalAct}) is just a special case of this more general $\tilde{\mathcal P}_{\phi}^{-1}$.

\noindent  We still impose the condition that $\tilde{\mathcal P}_{\phi}^{-1}(\vec{x})$ goes to its complex conjugate when $\tilde{V}(\vec{x})$ flips sign (along with the corresponding extra symmetry of the field $\psi$ as discussed before) to make the kinetic term real.

\noindent Let us assume $\tilde{\mathcal P}_{\phi}^{-1}$ can be expanded in terms of powers of $\partial_{\phi}^2$:
\begin{eqnarray}
\tilde{\mathcal P}_{\phi}^{-1}=b_0+b_2\partial_{\phi}^2+b_4\partial_{\phi}^4+\ldots,
\end{eqnarray}
where $b_i$ ($i=0, 2, 4, \ldots$) are function of the discrete metric data $\vec{x}$. So now we can start looking into the details with some concrete simple cases by truncating $\tilde{\mathcal P}_{\phi}^{-1}$ to finite orders. Notice that these truncations can also be seen (for appropriately chosen coefficients) as approximations of the GFT model defined in the previous section and corresponding to a standard relativistic scalar field. Such approximations amount to assuming a slow variation of the GFT field $\psi$ with respect to the variables $\phi$, and are quite natural from both an hydrodynamic perspective and from an effective field theory point of view.

\subsection{2nd order truncation}
Firstly we truncate $\tilde{\mathcal P}_{\phi}^{-1}$ to the second order:
\begin{eqnarray}
\tilde{S}_{\phi\rm G(2)}^{\rm GFT}&=&\frac{1}{2}\int {\psi}\star \tilde{{\mathcal P}}_{\phi\rm G(2)}^{-1}\star\psi+\int {\mathcal V}_{\phi\rm G}\star\left(\prod_{n=1}^{5}\psi\right)\quad +\quad \text{c.c.},
\end{eqnarray}
where
\begin{eqnarray}
{\mathcal V}_{\phi\rm G}&=& {\mathcal V}_{\phi} \star V_{\rm G}\\
\tilde{{\mathcal P}}_{\phi\rm G(2)}^{-1}&=&P_{\rm G}^{-1}\star \tilde{{\mathcal P}}_{\phi(2)}^{-1}\\
\tilde{\mathcal P}_{{\phi}(2)}^{-1}&=&b_2\partial_{\phi}^2+b_0.
\end{eqnarray}
For simplicity, the arguments of the GFT field, the kinetic kernel and the interaction kernel are not shown explicitly, as they remain the same as in $\tilde{S}_{\phi\rm G}^{\rm GFT}$. 

\noindent We compute ${\mathcal P}_{{\phi}(2)}$ by writing it in terms of the variable conjugate to $\phi$, i.e. the momentum of the discretized scalar field:
\begin{eqnarray}
{\mathcal P}_{{\phi}(2)}&=&\int_{-\infty}^{\infty} dp \frac{1}{-b_2p^2+b_0}e^{-i(\phi-\phi') p}.
\end{eqnarray}
As discussed around Eq. (\ref{tildeP_P}), the difference between ${\mathcal P}_{{\phi}(2)}$ and the exact ($\star$-product involved) inverse of $\tilde{\mathcal P}_{{\phi}(2)}$ is of the order of ${\rm O}(\hbar)$, so it does not affect the identification of the continuum counterpart of the discrete path integral corresponding to this action.

\noindent Here, $b_0$ is assumed to be non-zero, because we want our theory to reduce to the pure gravity GFT once we switch off the scalar field coupling. Using the contour integral, we have two cases
\begin{enumerate}
\item When ${\rm Im}(\sqrt{\frac{b_0}{b_2}})=0$,
\begin{eqnarray}
{\mathcal P}_{{\phi}(2)}=\frac{\pi}{\sqrt{b_0 b_2}}\sin(|\phi-\phi'|\sqrt{\frac{b_0}{b_2}}).
\end{eqnarray}
In order to identify the corresponding continuum theory, we expand the log of ${\mathcal P}_{{\phi}(2)}$ around $|\phi-\phi'|=0$,
\begin{eqnarray}
\ln \left({\mathcal P}_{{\phi}(2)}\right)=\ln\left(\frac{\pi}{b_2}\right)+\ln\left( |\phi-\phi'|\right) - \frac{b_0}{6 b_2}|\phi-\phi'|^2 + {\rm O}\left(|\phi-\phi'|^4\right).
\end{eqnarray}
The resulting ${\mathcal P}_{{\phi}(2)}$ is not smooth at $\phi-\phi'=0$, so the expansion should anyway be understood as the connection of the two expansion for $\phi-\phi'>0$ and $\phi-\phi'<0$ at the point $\phi-\phi'=0$. 
Still, the dominant term $\ln\left( |\phi-\phi'|\right)$ will blow up in the limit where $\phi=\phi'$, which is part of a naive continuum limit. This suggests that $\ln \left({\mathcal P}_{{\phi}(2)}\right)$ does not correspond to a discretization of the kinetic term of a proper continuum scalar field theory.

\item When ${\rm Im}(\sqrt{\frac{b_0}{b_2}})\neq0$,
\begin{eqnarray}
{\mathcal P}_{{\phi}(2)}=\frac{-i \pi}{\sqrt{b_2 b_0}}{\rm SI}(\sqrt{\frac{b_0}{b_2}})e^{i {\rm SI}(\sqrt{\frac{b_0}{b_2}})\sqrt{\frac{b_0}{b_2}(\phi-\phi')^2}},
\end{eqnarray}
where the function ${\rm SI}$ is defined as ${\rm SI}(z)\equiv{\rm Sgn}({\rm Im}(z)), \, z\in {\mathbb C}$.\\
One would naturally try to associate this propagator to the discretization of a continuum theory ${\mathbb K}^{(2)}$ defined as
\begin{eqnarray}
[dx^4]\sqrt{|{\rm g}|}{\mathbb K}^{(2)}&=&[dx^4]\sqrt{|{\rm g}|}\sqrt{\partial_\mu \phi \partial^\mu \phi}\nonumber\\
&\overset{\rm discretization}{\longrightarrow}& \tilde{V}\sqrt{\frac{(\phi-\phi')^2}{L^2}}+{\rm O}(a).
\end{eqnarray}
For this to hold, one needs
\begin{eqnarray}
i{\rm SI}(\sqrt{\frac{b_0}{b_2}})\sqrt{\frac{b_0}{b_2}}&=& \frac{i}{\hbar |L|} \tilde{V}\\
\frac{-i \pi}{\sqrt{b_2 b_0}}{\rm SI}(\sqrt{\frac{b_0}{b_2}})&=&1,
\end{eqnarray}
which gives
\begin{eqnarray}
b_0&=& -\frac{i \pi \tilde{V}}{\hbar |L|}\\
b_2&=& \frac{\pi \hbar |L|}{i  \tilde{V}}.
\end{eqnarray}
However, for the same relations we can see that ${\rm Im}(\sqrt{\frac{b_0}{b_2}})=0$, which contradicts our initial assumption. \end{enumerate}
In the end, it would seem that a GFT model defined by the above action, with only second derivatives in the scalar field variables, does not correspond to any reasonable discrete path integral for a scalar field.

\noindent This of course does not mean that the same GFT model is not well-defined or useful. It means however that we have no motivation from a discrete gravity point of view to treat it as physically sound as a fundamental model, and it should be rather treated as a useful approximation of the model defined in the previous section. In fact, this is also the approximation used in \cite{Oriti:2016qtz}, in the context of GFT condensate cosmology.

Still, one may find this result counterintuitive, exactly because $\tilde{\mathcal P}_{{\phi}(2)}^{-1}$ could be a good approximation of $\tilde{P}_{\phi}^{-1}$ when $\partial_{\phi}^2\ll 1$ and one may expect that it would correspond to some approximation of the standard scalar field kinetic term for ${\mathbb K}_{\phi}^{(2)}$. However this is not true, since it neglects the fact that in order to derive one from the other we have to compute an inverse function. More precisely, when we derive ${\mathcal P}_{{\phi}(2)}$ from $\tilde{\mathcal P}_{{\phi}(2)}^{-1}$, we need to integrate from $-\infty$ to $+\infty$ for $p$, which corresponds to $\partial_{\phi}$ in the Fourier transform. So the condition $\partial_{\phi}^2\ll 1$ does not hold during the deriving of ${\mathcal P}_{{\phi}(2)}$. In other words, there is no correspondent for ${\mathcal P}_{{\phi}(2)}$ of the condition $\partial_{\phi}^2\ll 1$.
\noindent From a more physical point of view, one can see the above as the reflection of the fact that $\tilde{\mathcal P}_{{\phi}}^{-1}$ is a classical object (part of the GFT equations of motion), while ${\mathcal P}_{{\phi}}$ corresponds to a quantum propagator, which actually depends from all $p$ contributions in $\tilde{\mathcal P}_{{\phi}}^{-1}$. There is no reason for us to expect a (small $p$) approximation on the classical level to have a correspondent at the quantum level.

\subsection{4th order truncation}
Now let us consider the next higher order truncation:
\begin{eqnarray}
\tilde{S}_{\phi\rm G(4)}^{\rm GFT}&=&\frac{1}{2}\int {\psi}\star \tilde{{\mathcal P}}_{\phi\rm G(4)}^{-1}\star\psi+\int {\mathcal V}_{\phi\rm G}\star\left(\prod_{n=1}^{5}\psi\right)\quad +\quad \text{c.c.},
\end{eqnarray}
where
\begin{eqnarray}
\tilde{{\mathcal P}}_{\phi\rm G(4)}^{-1}&=&P_{\rm G}^{-1}\star \tilde{{\mathcal P}}_{\phi(4)}^{-1}\\
\tilde{\mathcal P}_{{\phi}(4)}^{-1}&=&b_4\partial_{\phi}^4+b_2\partial_{\phi}^2+b_0.
\end{eqnarray}
Similarly to the $2$nd truncation case, a Fourier transformation can be performed on ${\mathcal P}_{{\phi}(4)}$
\begin{eqnarray}
{\mathcal P}_{{\phi}(4)}&=&\int_{-\infty}^{\infty} dp \frac{1}{b_4p^4-b_2p^2+b_0}e^{-i(\phi-\phi') p}.
\end{eqnarray}
We denote the roots of $b_4p^4-b_2p^2+b_0=0$ as $p=\pm r_1$, $p=\pm r_2$, where
\begin{eqnarray}
r_1&=&\frac{1}{\sqrt{2}}\sqrt{\frac{b_2}{b_4}-\sqrt{{(\frac{b_2}{b_4})}^2-4 \frac{ b_0}{b_4} }}\\
r_2&=&\frac{1}{\sqrt{2}}\sqrt{\frac{b_2}{b_4}+\sqrt{{(\frac{b_2}{b_4})}^2-4 \frac{ b_0}{b_4} }}.
\end{eqnarray}
The expression of ${\mathcal P}_{{\phi}(4)}$ depends on whether $r_1$ or $r_2$ is real or complex, so we give ${\mathcal P}_{{\phi}(4)}$ in the following cases:
\begin{itemize}
\item $r_1 \neq r_2$ case:
\begin{enumerate}
\item When ${\rm Im}(r_1)\neq 0$ and ${\rm Im}(r_2)\neq 0$, we have
\begin{eqnarray}
{\mathcal P}_{{\phi}(4)}= \frac{i \pi}{R_1(R_1^2-R_2^2)b_4}e^{iR_1|\phi-\phi'|} + \frac{i \pi }{R_2(R_2^2-R_1^2)b_4}e^{iR_2|\phi-\phi'|},
\end{eqnarray}
where $R_1={\rm SI}(r_1)r_1$ and $R_2={\rm SI}(r_2)r_2$. Then as in the ${\mathcal P}_{{\phi}(2)}$ case, we expand
\begin{eqnarray}\label{FourthExpan}
\ln \left({\mathcal P}_{{\phi}(4)}\right)&=&\ln\left(-\frac{i\pi}{R_1R_2(R_1+R_2)b_4}\right) + \frac{1}{2}R_1R_2|\phi-\phi'|^2 + \nonumber\\
&&\frac{i}{6}R_1R_2(R_1+R_2)|\phi-\phi'|^3 + {\rm O}\left(|\phi-\phi'|^4\right).
\end{eqnarray}
Now we need to find the corresponding propagating term ${\mathbb K}^{(4)}$ for the continuum theory. Notice that in Eq. (\ref{FourthExpan}), since all the terms must be of the same dimension (dimensionless actually), in which case the overall order of a term can be indicated by the order of $|\phi-\phi'|$, the lowest order term that involves the scalar field is the second order $\frac{1}{2}R_1R_2|\phi-\phi'|^2$, which we will assume is the dominant term in the continuum limit. So the same as in the previous discussion, we get
\begin{eqnarray}
{\mathbb K}^{(4)}=\partial_\mu \phi \partial^\mu \phi,
\end{eqnarray}
if
\begin{equation}
\frac{1}{2}R_1R_2 = \frac{i \tilde{V}}{\hbar L^2}
\end{equation}
\begin{equation}
-\frac{i\pi}{R_1R_2(R_1+R_2)b_4} = 1,
\end{equation}
which, after reading the dimension of $b_i$ from Eq. (\ref{NormalScalAct}), leads to
\begin{eqnarray}
b_4&=&\alpha\left(\frac{i\hbar L^2}{ \tilde{V}}\right)^{\frac{3}{2}}\label{b4}\\
b_2&=&\left(4\alpha-\frac{\pi^2}{4\alpha}\right)\left(\frac{i\hbar L^2}{ \tilde{V}}\right)^{\frac{1}{2}}\label{b2}\\
b_0&=&4\alpha\left(\frac{i\hbar L^2}{ \tilde{V}}\right)^{-\frac{1}{2}}\label{b0}
\end{eqnarray}
where $\alpha$ is a dimensionless constant and $\alpha\neq \pm \frac{\pi}{4\sqrt{2}}$ to make sure $r_1 \neq r_2$. One also needs to make sure that ${\rm Im}(r_1)\neq 0$ and ${\rm Im}(r_2)\neq 0$.  At last by requiring $\alpha$ being real, one can make sure ${\mathcal P}_{{\phi}(4)}$ goes to the complex conjugate when $\tilde{V}$ flips sign. Then using Eq. (\ref{VLV}), we can write $b_4$, $b_2$, $b_0$ explicitly as functions of $\vec{x}$. Actually, we need the above relations to be satisfied only in a semi-classical limit, but we will refrain from searching for more general functions with the same limit.

Now with the help of Eq. (\ref{b4}, \ref{b2}, \ref{b0}), we find all the terms in Eq. (\ref{FourthExpan}) as proportional to $\left( \tilde{V}\frac{|\phi-\phi|^2}{ L^2}\right)^{w}, \,\,w=1,\frac{3}{2},2,\frac{5}{2},\ldots$, which means the dominant order in a continuum approximation of Eq. (\ref{FourthExpan}) is indeed $|\phi-\phi'|^2$.
\item When one of ${\rm Im}(r_1)$ and ${\rm Im}(r_2)$ equals to zero (let us choose ${\rm Im}(r_1)=0$ and ${\rm Im}(r_2)\neq0$ without loss of generality), we can go through all the calculations of the previous case, to find
\begin{eqnarray}
r_1\sim\tilde{V}^{\frac{1}{2}}.
\end{eqnarray}
However, since we integrate over both orientations in the Feynman amplitude, which means that $\tilde{V}$ flips sign in the integral, the condition ${\rm Im}(r_1)=0$ does not hold. So we do not have a consistent solution in this case. In any case, even leaving this problem aside, we would still get the same ${\mathbb K}^{(4)}$ as in the previous case, with $b_4$, $b_2$, $b_0$ only differing by constant factors.
\item When ${\rm Im}(r_1)={\rm Im}(r_2)=0$, we have
\begin{eqnarray}
{\mathcal P}_{{\phi}(4)}=\frac{\pi}{r_1(r_2^2-r_1^2)b_4}\sin(r_1|\phi-\phi'|)+\frac{\pi}{r_2( r_1^2-r_2^2)b_4}\sin(r_2|\phi-\phi'|),
\end{eqnarray}
and
\begin{eqnarray}\label{R1realR2real}
\ln\left({\mathcal P}_{{\phi}(4)}\right)= \ln\left(\frac{\pi}{6b_4}\right)+3\ln\left(|\phi-\phi'|\right)-\frac{r_1^2+r_2^2}{20}|\phi-\phi'|^2+{\rm O}\left(|\phi-\phi'|^4\right),
\end{eqnarray}
which does not relate to any continuum theory for the same reason as in the case (1) of ${\mathcal P}_{{\phi}(2)}$.
\end{enumerate}
\item $r_1 = r_2$ case: from $r_1 = r_2$, we have $b_2^2=4b_0b_4$ and $r \equiv r_1 = r_2=\sqrt{\frac{b_2}{2b_4}} $.
\begin{enumerate}
\item When ${\rm Im}(r)\neq 0$, we have
\begin{eqnarray}
{\mathcal P}_{{\phi}(4)}=-\frac{i\pi}{2 R^3 b_4}e^{iR|\phi-\phi'|}-\frac{\pi}{2R^2 b_4}|\phi-\phi'|e^{iR|\phi-\phi'|}
\end{eqnarray}
and
\begin{eqnarray}
\ln \left({\mathcal P}_{{\phi}(4)}\right) = \ln\left(-\frac{i\pi }{2R^3b_4}\right) + \frac{R^2}{2}|\phi-\phi'|^2 + \frac{i R^3}{3} |\phi-\phi'|^3  + {\rm O}\left(|\phi-\phi'|^{4}\right),
\end{eqnarray}
where $R={\rm SI}(r)r$. So a corresponding continuum kinetic term can be identified as ${\mathbb K}^{(4)}=\partial_\mu \phi \partial^\mu \phi$, if
\begin{eqnarray}
\frac{R^2}{2}&=&\frac{i \tilde{V}}{\hbar L^2}\\
-\frac{i\pi }{2R^3b_4}&=&1.
\end{eqnarray}
In the end, we have
\begin{eqnarray}
b_4&=&\frac{\pi}{4\sqrt{2}}\left(\frac{i\hbar L^2}{ \tilde{V}}\right)^{\frac{3}{2}}\\
b_2&=&\pm\frac{\pi}{\sqrt{2}}\left(\frac{i\hbar L^2}{ \tilde{V}}\right)^{\frac{1}{2}}\\
b_0&=&\frac{\pi}{\sqrt{2}}\left(\frac{i\hbar L^2}{ \tilde{V}}\right)^{-\frac{1}{2}},
\end{eqnarray}
which is just a special case of the case (1) of the $r_1 \neq r_2$ case when $\alpha=\pm\frac{\pi}{4\sqrt{2}}$, which is also easily checked to be consistent with the condition ${\rm Im}(r)\neq 0$ and ${\mathcal P}_{{\phi}(4)}$ goes to the complex conjugate when $\tilde{V}$ flips sign.
\item When ${\rm Im}(r)=0$, we have
\begin{eqnarray}
{\mathcal P}_{{\phi}(4)}=\frac{\pi}{2R^3b_4}\sin( R|\phi-\phi'|)-\frac{\pi}{2R^2b_4} |\phi-\phi'|\cos( R|\phi-\phi'|)
\end{eqnarray}
and
\begin{eqnarray}
\ln \left({\mathcal P}_{{\phi}(4)}\right)=\ln\left(\frac{\pi}{6b_4}\right)+3\ln\left(|\phi-\phi'|\right)-\frac{R^2}{10}|\phi-\phi'|^2+{\rm O}\left(|\phi-\phi'|^4\right).
\end{eqnarray}
So for the same reason as in the case (1) of $\tilde{\mathcal P}_{{\phi}(2)}^{-1}$, this case can not be related to a nice continuum theory.
\end{enumerate}
\end{itemize}

To summarise, the 4th order truncation $\tilde{\mathcal P}_{{\phi}(4)}^{-1}$ corresponds to a continuum kinetic term ${\mathbb K}^{(4)}=\partial_\mu \phi \partial^\mu \phi$, if the coefficients $b_4$, $b_2$, $b_0$ are chosen as in Eq. (\ref{b4}, \ref{b2}, \ref{b0}), where $\alpha$ is a arbitrary constant as long as ${\rm Im}(r_1)\neq 0$ and ${\rm Im}(r_2)\neq 0$. Other cases do not seem to relate to a consistent continuum theory.

\subsection{Higher order truncations}
Let us also see if we can get some insight into the more general, higher order truncations. We consider a truncation to $2N$th order: $\tilde{\mathcal P}_{{\phi}(2N)}^{-1}$, $N=2, 3, 4,\ldots$, for which we denote the roots of $\tilde{\mathcal P}_{{\phi}(2N)}^{-1}(-p^2)=0$ as $p=\pm r_i$ ($i=1,2,\ldots,N$). \\
\noindent We also assume ${\rm Im}(r_i)\neq 0$, which seems to be reasonable, since from our discussion for $\tilde{\mathcal P}_{{\phi}(4)}^{-1}$, we see the condition ${\rm Im}(r_i)= 0$ is impossible to hold in the Feynman amplitude if we integrate both orientations of the volume element.\\
\noindent Then, in this case, we find that $\ln ({\mathcal P}_{{\phi}(2N)} )$ can still be expanded as an infinite polynomial of $|\phi-\phi'|$:
\begin{eqnarray}
\ln\left({\mathcal P}_{{\phi}(2N)}\right)=\sum_{m=0}^{\infty}c_m|\phi-\phi'|^m,
\end{eqnarray}
but the coefficients $c_m$ vanish for $m=1, 3, \ldots, 2N-3$. In other words, adding higher order terms in $\tilde{\mathcal P}_{{\phi}(2N)}^{-1}$, we kill the lowest odd order terms in ${\mathcal P}_{{\phi}(2N)}$ one by one. We regain a smooth function when $N\rightarrow\infty$, i.e. for $\tilde{P}_{\phi}^{-1}$. However, the leading order is always the second order, which means we only reproduce ${\mathbb K}^{(2N)}=\partial_\mu \phi \partial^\mu \phi$ for the continuum scalar field theory, at leading order, also from this generalised finite truncation\footnote{In the appendix \ref{app:truncation}, we partially prove this result: we explicitly prove this if all $r_i$ are different (e.g. all $r_i$ are first order roots); but if there are higher order roots, we only have a conjecture that has already been tested to be true for several simple cases, and it suggests this results is still valid (where we can view $\tilde{\mathcal P}_{{\phi}(4)}^{-1}$ as a fully discussed example).}.\\

\subsection{The minimal coupling}
Now we summarize what has been learned from the above discussion. For $\tilde{\mathcal P}_{\phi}^{-1}$ a finite polynomial of $\partial_{\phi}^2$ to the $2N$th order:
\begin{eqnarray}
\tilde{S}_{\phi\rm G(2N)}^{\rm GFT}&=&\frac{1}{2}\int {\psi}\star \tilde{{\mathcal P}}_{\phi\rm G(2N)}^{-1}\star\psi+\int {\mathcal V}_{\phi\rm G}\star\left(\prod_{n=1}^{5}\psi\right)\quad +\quad \text{c.c.},
\end{eqnarray}
where
\begin{eqnarray}
\tilde{{\mathcal P}}_{\phi\rm G(4)}^{-1}&=&P_{\rm G}^{-1}\star \tilde{{\mathcal P}}_{\phi(2N)}^{-1}\\
\tilde{\mathcal P}_{{\phi}(2N)}^{-1}&=&\sum_{n=0}^{N}b_{2n}\partial_{\phi}^{2n} \qquad ,
\end{eqnarray}
we have as a leading order continuum theory ${\mathbb K}^{(2N)}=\partial_\mu \phi \partial^\mu \phi$ for $N>1$. When $N=1$ no continuum theory can be related consistently to the corresponding discrete path integral.\\
\noindent As for the infinite polynomials ($N\rightarrow \infty$), we only discuss (in a "reverse engineering" way, in the previous section) the special case $\tilde{\mathcal P}_{\phi}^{-1}\sim e^{b \partial_{\phi}^2}$, where $b$ is a function of $\vec{x}$, and we still have ${\mathbb K}=\partial_\mu \phi \partial^\mu \phi$ for the continuum scalar field kinetic term.
However this is the context where more general forms of $\tilde{\mathcal P}_{\phi}^{-1}$ can be considered, leading to more possibilities for the continuum theory. In fact, considering actions of the type ${\mathbb K}=\left(\partial_\mu \phi \partial^\mu \phi\right)^{\frac{3}{2}}, \left(\partial_\mu \phi \partial^\mu \phi\right)^{\frac{4}{2}}, \left(\partial_\mu \phi \partial^\mu \phi\right)^{\frac{5}{2}}$, and setting up the  the same "reverse engineering" procedure we applied for the ${\mathbb K}=\partial_\mu \phi \partial^\mu \phi$ case, we can find the corresponding $\tilde{\mathcal P}_{\phi}^{-1}$ as complicated functions involving hypergeometric functions and gamma functions, but that can still be expanded as infinite polynomials of $\partial_{\phi}^2$.
We do not present in detail these other cases, but it is important to stress that, in principle, any scalar field dynamics encoded in the kinetic term as a function of $\partial_\mu \phi \partial^\mu \phi$ as ${\mathbb K}(\partial_\mu \phi \partial^\mu \phi)$ can be handled in our framework. In other words, we have found a general way to include the minimal coupling of a scalar field within the GFT framework.

\section{Generalizations and applications}\label{sec:ga}
Let us now discuss some possible generalizations of our construction and results. We will also outline the application of our results in the context of GFT condensate cosmology, for what concers the effective cosmological dynamics they originate and the definition of relational observables.

\subsection{Multiple (complex) scalar field coupling}
It is straightforward to generalize our model to the coupling of multiple scalar fields, which in particular includes the case of complex scalar fields. There is no limitation, in fact, for the number of independent degrees of freedom that we can handle per point, at each vertex of the dual complex $\Delta*$. \\
\noindent Consider $M$ real scalar fields and looking for the same minimal coupling, we simply add $M$ new arguments in the GFT field as
\begin{eqnarray}
\psi^{\rm M}=\psi^{\rm M}(\vec{x}; \vec{\phi}),
\end{eqnarray}
where $\vec{\phi}=(\phi^{(1)}, \ldots, \phi^{(M)})\in {\mathbb R}^M$ simply denoting the $M$ real scalar fields.\\
\noindent Then we generalize $\tilde{\mathcal P}_{{\phi }}^{-1}$ to include the $M$ real scalar field propagation:
\begin{eqnarray}
\tilde{\mathcal P}_{{\phi }}^{-1}\rightarrow\tilde{\mathcal P}_{{\phi \rm M}}^{-1}(\vec{x}; \partial^2_{\vec{\phi}})\equiv\tilde{\mathcal P}_{{\phi }(1)}^{-1}(\vec{x}; \partial^2_{\phi^{(1)}})\star\ldots\star\tilde{\mathcal P}_{{\phi }(M)}^{-1}(\vec{x}; \partial^2_{\phi^{(M)}}),
\end{eqnarray}
where $\tilde{\mathcal P}_{{\phi }(m)}^{-1}$ denotes a general function of $\vec{x}$ and $\partial^2_{\phi^{(m)}}, m\in\{1,\ldots,M\}$, and one has to pay special care to the ordering of the $\star$-products, as discussed in the previous sections.\\
\noindent As for the scalar field potential, we generalize ${\mathcal V}_{\phi }$ as
\begin{eqnarray}
{\mathcal V}_{\phi }\rightarrow{\mathcal V}_{\phi \rm M}={\mathcal V}_{\phi \rm M}(\vec{x}_1,\ldots,\vec{x}_5;\vec{\phi}).
\end{eqnarray}
\noindent The GFT action incorporating the minimal coupling of the $M$ real scalar fields is then
\begin{eqnarray}
\tilde{S}_{\phi \rm MG}^{\rm GFT}&=&\frac{1}{2}\int [dx^4][dx'^4][d{\phi}^M] {\psi}^{\rm M}(\vec{x};\vec{\phi})\star \tilde{{\mathcal P}}_{\phi \rm MG}^{-1}(\vec{x},\vec{x}';\partial^2_{\vec{\phi}})\star\psi^{\rm M}(\vec{x}';\vec{\phi})+\nonumber\\
&&\int[d{\phi}^M]\left(\prod_{n=1}^{5}[dx_n^4]\right){\mathcal V}_{\phi \rm MG}(\vec{x}_1,\ldots,\vec{x}_5;\vec{\phi})\star\left(\prod_{n=1}^{5}\psi^{\rm M}(\vec{x}_n;\vec{\phi})\right)\quad +\quad \text{c.c.},
\end{eqnarray}
where
\begin{eqnarray}
\tilde{{\mathcal P}}_{\phi \rm MG}^{-1}(\vec{x},\vec{x}';\partial^2_{\vec{\phi}})&=&P_{\rm G}^{-1}(\vec{x},\vec{x}')\star \tilde{{\mathcal P}}_{\phi \rm M}^{-1}(\vec{x}';\partial^2_{\vec{\phi}})\\
{\mathcal V}_{\phi \rm MG}(\vec{x}_1,\ldots,\vec{x}_5;\vec{\phi})&=&{\mathcal V}_{\phi \rm M}(\vec{x}_1,\ldots,\vec{x}_5;\vec{\phi}) \star V_{\rm G}(\vec{x}_1,\ldots,\vec{x}_5).
\end{eqnarray}
\noindent Then we have the propagator $\tilde{{\mathcal P}}_{\phi \rm MG}$ as
\begin{eqnarray}
\tilde{{\mathcal P}}_{\phi \rm MG}(\vec{x},\vec{x}';\vec{\phi},\vec{\phi}')&=&\tilde{{\mathcal P}}_{\phi \rm M}(\vec{x};\vec{\phi},\vec{\phi}')\star P_{\rm G}(\vec{x},\vec{x}')\\
\tilde{{\mathcal P}}_{\phi \rm M}(\vec{x};\vec{\phi},\vec{\phi}')&=&\tilde{\mathcal P}_{{\phi }(M)}\left(\vec{x};\left(\phi'^{(M)}-\phi^{(M)}\right)^2\right)\star\ldots\star\tilde{\mathcal P}_{{\phi }(1)}\left(\vec{x};\left(\phi'^{(1)}-\phi^{(1)}\right)^2\right)
\end{eqnarray}
with
\begin{eqnarray}
\tilde{\mathcal P}_{{\phi }(m)}^{-1}(\vec{x}; \partial^2_{\phi^{(m)}})\star\tilde{\mathcal P}_{{\phi }(m)}\left(\vec{x};\left(\phi'^{(M)}-\phi^{(M)}\right)^2\right)=\delta\left(\phi^{(m)},\phi'^{(m)}\right),
\end{eqnarray}
so that
\begin{eqnarray}
\int[dx'^4]\tilde{{\mathcal P}}_{\phi \rm MG}^{-1}(\vec{x},\vec{x}';\partial^2_{\vec{\phi}})\star\tilde{{\mathcal P}}_{\phi \rm MG}(\vec{x}',\vec{x}'';\vec{\phi},\vec{\phi}')=\left(\prod_{n=1}^4\delta^\star_{x_n}(x''_n)\right)\left(\prod_{m=1}^M\delta\left(\phi^{(m)},\phi'^{(m)}\right)\right) \quad . \nonumber
\end{eqnarray}
\noindent Thus, given a Feynman diagram (dual to) $\Delta$, its Feynman amplitude gives rise to the discrete path integral for gravity minimally coupled to $M$ real scalar fields, following the same steps outlined in the previous sections:
\begin{eqnarray}
\int_{\Delta}\prod_{\vec{\star}}\tilde{{\mathcal P}}_{\phi \rm MG}{\mathcal V}_{\phi \rm MG} \,=\,\int \prod_{f\in \Delta}dB^{\rm C}_f \prod_{l\in \Delta*} dg_l \; \left(\prod_{m=1}^M\prod_{v\in\Delta} d\phi_{v}^{(m)}\right)z^{\Delta}_{\phi \rm M}z^{\Delta}_{\rm G},
\end{eqnarray}
where
\begin{eqnarray}
z^{\Delta}_{\phi \rm M}=(1+{\rm O}(\hbar))e^{\frac{i}{\hbar}\left(\sum_{m=1}^M\sum_{l\in \Delta}\tilde{V}{\mathbb K}_{(m)}^{\Delta}\left(\left(\frac{\delta_l \phi^{(m)}}{L}\right)^2\right)+\frac{i}{\hbar}\sum_{v\in \Delta}V{\mathbb V}_{\rm M}^{\Delta}(\vec{\phi}_{v})\right)},
\end{eqnarray}
and
\begin{eqnarray}
{\mathbb K}_{(m)}^{\Delta}\left(\left(\frac{\phi'^{(m)}-\phi^{(m)}}{L}\right)^2\right)&=&\frac{\hbar}{ i \tilde{V}}\ln\left(\tilde{\mathcal P}_{{\phi }(m)}\left(\vec{x};\left(\phi'^{(m)}-\phi^{(m)}\right)^2\right)\right)\\
{\mathbb V}_{\rm M}^{\Delta}(\vec{\phi})&=&\frac{\hbar}{ i V}\ln\left({\mathcal V}_{\phi \rm M}(\vec{x}_1,\ldots,\vec{x}_5; \vec{\phi})\right).
\end{eqnarray}

\subsection{Effective cosmological dynamics and relational observables}
As anticipated, our construction of a coupled GFT model for gravity and a scalar field has centered on those features that are going to survive a continuum approximation, even though it has been based on the analysis of the discrete path integral arising as the Feynman amplitude of the same GFT model. Our focus on the GFT action, rather than the Feynman amplitudes themselves or their equivalent spin foam expression, is also motivated by the fact that the GFT formalism offers new ways to extract effective continuum physics from both discrete path integrals and spin foam models.

One strategy to do so has been developed in a series of papers and cocnerned in particular the effective cosmological dynamics that can be extracted from GFT (thus spin foam) models. We refer to it as GFT condensate cosmology, since it based on condensate states in the GFT formalism \cite{Gielen2014a,PhysRevLett.111.031301,Gielen:2016dss}. In this context, the minimal coupling of a real scalar field had been considered \cite{Oriti:2016qtz,Oriti:2016ueo,Cesare2017,PhysRevD.94.064051,PhysRevD.95.064004} and played an instrumental role in the extraction of a cosmological dynamics with the correct semiclassical limit, which allowed also an explicit link with loop quantum cosmology \cite{Ashtekar2011,Banerjee:2011qu,lrr-2008-4}.

Our coupled model can be seen as providing a more solid basis for the analysis performed in these papers, since it results in a model with the same features that were {\it assumed} there. In particular, just like in those works, our model incorporates a real scalar field by extending the domain of the GFT field $\varphi(\vec{x})\rightarrow\psi(\vec{x};\phi)$ and it is local in the new variable. In fact, the scalar field variable enters the GFT action in such a way that it could be treated (in the appropriate approximation) like a time variable, and it is indeed used as a relational time in the effective cosmological dynamics and in the definition of (relational) observables of the theory. Moreover, in the simplest approximation of our coupled GFT model, where the kinetic term is truncated to the second order in the derivatives with respect to the scalar field variable, it coincides with the one used in \cite{Oriti:2016qtz,Oriti:2016ueo}.

We expect that a generalization of our model to multiple scalar fields can be used, in a similar vein, to construct relational observables of a more refined type and to study local physics in a fully diffeomorphism invariant language, in the spirit of the dust frame in cosmology or Brown-Kuchar model \cite{Brown:1994py}, going beyond the spatially homogenous setting.

\

\noindent According to the GFT condensate cosmology, for the simplest type of GFT condensate states (corresponding to a mean field approximation of the full GFT path integral), the classical GFT equations of motion provide a cosmological dynamics for a homogeneous universe in the sense of a non-linear extension of quantum cosmology:
\begin{eqnarray}
\tilde{\mathcal P}^{-1}_{\phi\rm G}\psi+5\left(\int {\mathcal V}_{\phi\rm G}{\psi}^4\right)=0,
\end{eqnarray}
where $\psi$ plays the role of the condensate wave function for the quantum universe, but should be understood as an hydrodynamic variable, like in real Bose condensates, where it encodes both the density and the velocity of the fluid. \\
\noindent In the approximation in which the GFT interactions are subdominant, which corresponds to the regime in which the discrete gravity path integral and the dual spin foam amplitudes are relevant,  we have the free equation:
\begin{eqnarray}
\tilde{\mathcal P}^{-1}_{\phi\rm G}\psi=\tilde{\mathcal P}^{-1}_{\phi\rm G}(\vec{x},\vec{x}';\partial^2_\phi)\psi=0 \qquad .
\end{eqnarray}
\noindent If we assume that $\tilde{\mathcal P}^{-1}_{\phi\rm G}$ can be expanded for $\partial_{\phi}^2$, we have
\begin{eqnarray}
\tilde{\mathcal P}^{-1}_{\phi\rm G}\psi=\left(\tilde{b}_0+\tilde{b}_2\partial_{\phi}^2+\tilde{b}_4\partial_{\phi}^4+\ldots\right)\psi=0,
\end{eqnarray}
where $\tilde{b}_0, \tilde{b}_2, \tilde{b}_4,\ldots$ are functions of $\vec{x}$ and $\vec{x}'$.\\

\noindent This type of equation can be compared with the dynamics for a homogeneous universe containing a single real scalar field (in addition to the metric) in quantum cosmology. There, the scalar field part of the Hamiltonian constraint is
\begin{eqnarray}
\hat{{\mathcal H}}_{\rm matter}=\frac{1}{4\sqrt{{\rm h}}}\partial_\phi^2-\sqrt{{\rm h}}\frac{\Lambda }{2\kappa},
\end{eqnarray}
where ${\rm h}=\det({\rm h}_{ij})$ is the determinant of spacial metric ${\rm h}_{ij}$, $\Lambda$ is the cosmological constant.\\
\noindent As mentioned before, we can always impose the closure constraint in $V_{\rm G}$ instead of $P_{\rm G}$ with the total convolution of the Feynman amplitude staying unchanged. Then we have
\begin{eqnarray}
\tilde{\mathcal P}^{-1}_{\phi\rm G}=P_{\rm G}^{-1}\star \tilde{\mathcal P}_{\phi }^{-1}\psi=\tilde{\mathcal P}_{\phi }^{-1}\psi=0.
\end{eqnarray}
\noindent Then the above expansion is simply the expansion of the part of the GFT kinetic term depending on the scalar field variable, which for our model reads:
\begin{eqnarray}
\tilde{\mathcal P}_{\phi }^{-1}(\partial_{\phi_v}^2)\psi=\left(\alpha_0\sqrt{\tilde{b}}+\frac{\alpha_2}{\sqrt{\tilde{b}}}\partial_{\phi}^2+\alpha_4\tilde{b}^{-\frac{3}{2}}\partial_{\phi}^4+\ldots\right)\psi=0,
\end{eqnarray}
where $\alpha_0, \alpha_2, \alpha_4,\ldots$ are constants and $\tilde{b}=\frac{\tilde{V}}{L^2}$. \\
\noindent This implies that we obtain an effective cosmological equation analogous to the quantum cosmological one, as a 2nd order truncation of our model in the derivatives with respect to the scalar field variable (as we have mentioned, this is natural in such hydordynamics setting and was indeed the strategy followed in \cite{Oriti:2016qtz,Oriti:2016ueo}). The coupling of the scalar field as well as the constant term match the ones of quantum cosmology in the approximation in which the geometric quantities $\tilde{b}$ are assumed to be just functions of the $3$-volume of the tetrahedron to which the GFT propagator is associated (in the GFT perturbative expansion) \cite{Calcagni:2012cv}.

\section{Conclusion}\label{sec:con}
\noindent We have shown how to introduce scalar field degrees of freedom in GFT fields and states, and construct a corresponding GFT model, in such a way that the corresponding Feynman amplitudes take the form of a simplicial path integral for gravity minimally coupled to a relativistic scalar field, with standard propagator and arbitrary interactions, and with the correct classical and continuum limit (in the sense of kinematical approximations, before considering the actual quantum dynamics of the GFT model). Implicitly, this defines also a coupled spin foam model. This extends the pure gravity case which also gives, in the same sense, a correct discrete path integral for simplicial gravity.  Some variations of the same GFT model have also been considered, and the corresponding simplicial path integrals investigated. Finally, we have discussed the straightforward generalization of our construction to multiple minimally coupled scalar fields, and also provided a quick comparison between the effective cosmological dynamics that would correspond to the mean field approximation of the quantum dynamics of GFT condensates in our coupled model and the traditional quantum cosmological dynamics for the same degrees of freedom.

\noindent Indeed, it is in this direction of the cosmology emergent from full quantum gravity that we expect our newly defined GFT model to find immediate interesting applications, along the lines of .\cite{Oriti:2016qtz,Oriti:2016ueo}. At the level of cosmological background dynamics, i.e. concerning homogeneous degrees of freedom only, the coupling of matter fields to gravity is already crucial for describing the correct physics of the early universe (e.g. via inflationary-type scenarios), including the dynamics of anisotropies, but could also be crucial for unraveling the true physics replacing the cosmological singularity. Moreover, our construction, generalized to the case of multiple scalar fields, would be the natural starting point for modeling dust matter in a GFT/spin foam context and for defining the physical frame in which to describe cosmological perturbations and, more generally, effective local continuum physics emerging from full quantum gravity. This would mean truly bridging the gap between Planck-scale physics and the effective physics at macroscopic scales.

\acknowledgments

YL is supported by China Scholarship Council (CSC) No.201506040138, NSFC (grant nos. 11475023 and 11235003), and the doctoral program of higher education of China. MZ acknowledges the funding received from Alexander von Humboldt Foundation. MZ and YL would like to thank Dr. Song He for the inspiring discussions.

\bibliographystyle{jhep}
\bibliography{GFTcpSF}

\providecommand{\href}[2]{#2}\begingroup\raggedright\begin{thebibliography}{10}

\bibitem{Oriti:2011jm}
D.~Oriti, {\it {The microscopic dynamics of quantum space as a group field
  theory}},  in {\em {Proceedings, Foundations of Space and Time: Reflections
  on Quantum Gravity: Cape Town, South Africa}}, pp.~257--320, 2011.
\newblock \href{http://arxiv.org/abs/1110.5606}{{\tt arXiv:1110.5606}}.

\bibitem{Krajewski:2012aw}
T.~Krajewski, {\it {Group field theories}},  {\em PoS} {\bf QGQGS2011} (2011)
  005, [\href{http://arxiv.org/abs/1210.6257}{{\tt arXiv:1210.6257}}].

\bibitem{Oriti:2014uga}
D.~Oriti, {\it {Group Field Theory and Loop Quantum Gravity}},  2014.
\newblock \href{http://arxiv.org/abs/1408.7112}{{\tt arXiv:1408.7112}}.

\bibitem{Oriti2015}
D.~Oriti, J.~P. Ryan, and J.~Th\"{u}rigen, {\it Group field theories for all
  loop quantum gravity},  {\em New Journal of Physics} {\bf 17} (2015), no.~2
  023042, [\href{http://arxiv.org/abs/1409.3150}{{\tt arXiv:1409.3150}}].

\bibitem{Ashtekar2004}
A.~Ashtekar and J.~Lewandowski, {\it Background independent quantum gravity: a
  status report},  {\em Classical and Quantum Gravity} {\bf 21} (2004), no.~15
  R53, [\href{http://arxiv.org/abs/gr-qc/0404018}{{\tt gr-qc/0404018}}].

\bibitem{Rovelli2004}
C.~Rovelli, {\em Quantum Gravity}.
\newblock Cambridge University Press, Cambridge University Press University
  Printing House Shaftesbury Road Cambridge CB2 8BS United Kingdom, 2004.

\bibitem{Thiemann2008}
T.~Thiemann, {\em Modern Canonical Quantum General Relativity}.
\newblock Cambridge University Press, Cambridge University Press University
  Printing House Shaftesbury Road Cambridge CB2 8BS United Kingdom, 2008.

\bibitem{MUXIN2007}
M.~HAN, Y.~MA, and W.~HUANG, {\it Fundamental structure of loop quantum
  gravity},  {\em International Journal of Modern Physics D} {\bf 16} (2007),
  no.~09 1397--1474, [\href{http://arxiv.org/abs/gr-qc/0509064}{{\tt
  gr-qc/0509064}}].

\bibitem{Oriti:2013aqa}
D.~Oriti, {\it {Group field theory as the 2nd quantization of Loop Quantum
  Gravity}},  {\em Class. Quant. Grav.} {\bf 33} (2016), no.~8 085005,
  [\href{http://arxiv.org/abs/1310.7786}{{\tt arXiv:1310.7786}}].

\bibitem{Alexandrov:2011ab}
S.~Alexandrov, M.~Geiller, and K.~Noui, {\it {Spin Foams and Canonical
  Quantization}},  {\em SIGMA} {\bf 8} (2012) 055,
  [\href{http://arxiv.org/abs/1112.1961}{{\tt arXiv:1112.1961}}].

\bibitem{Perez:2012wv}
A.~Perez, {\it {The Spin Foam Approach to Quantum Gravity}},  {\em Living Rev.
  Rel.} {\bf 16} (2013) 3, [\href{http://arxiv.org/abs/1205.2019}{{\tt
  arXiv:1205.2019}}].

\bibitem{Ambjorn:2012jv}
J.~Ambjorn, A.~Goerlich, J.~Jurkiewicz, and R.~Loll, {\it {Nonperturbative
  Quantum Gravity}},  {\em Phys. Rept.} {\bf 519} (2012) 127--210,
  [\href{http://arxiv.org/abs/1203.3591}{{\tt arXiv:1203.3591}}].

\bibitem{Rivasseau:2016rgt}
V.~Rivasseau, {\it {Constructive Tensor Field Theory}},  {\em SIGMA} {\bf 12}
  (2016) 085, [\href{http://arxiv.org/abs/1603.07312}{{\tt arXiv:1603.07312}}].

\bibitem{Carrozza:2016vsq}
S.~Carrozza, {\it {Flowing in Group Field Theory Space: a Review}},  {\em
  SIGMA} {\bf 12} (2016) 070, [\href{http://arxiv.org/abs/1603.01902}{{\tt
  arXiv:1603.01902}}].

\bibitem{Benedetti:2014qsa}
D.~Benedetti, J.~Ben~Geloun, and D.~Oriti, {\it {Functional Renormalisation
  Group Approach for Tensorial Group Field Theory: a Rank-3 Model}},  {\em
  JHEP} {\bf 03} (2015) 084, [\href{http://arxiv.org/abs/1411.3180}{{\tt
  arXiv:1411.3180}}].

\bibitem{Geloun:2016qyb}
J.~Ben~Geloun, R.~Martini, and D.~Oriti, {\it {Functional Renormalisation Group
  analysis of Tensorial Group Field Theories on $\mathbb{R}^d$}},  {\em Phys.
  Rev.} {\bf D94} (2016), no.~2 024017,
  [\href{http://arxiv.org/abs/1601.08211}{{\tt arXiv:1601.08211}}].

\bibitem{Carrozza:2016tih}
S.~Carrozza and V.~Lahoche, {\it {Asymptotic safety in three-dimensional SU(2)
  Group Field Theory: evidences in the local potential approximation}},
  \href{http://arxiv.org/abs/1612.02452}{{\tt arXiv:1612.02452}}.

\bibitem{Gielen:2016dss}
S.~Gielen and L.~Sindoni, {\it {Quantum Cosmology from Group Field Theory
  Condensates: a Review}},  {\em SIGMA} {\bf 12} (2016) 082,
  [\href{http://arxiv.org/abs/1602.08104}{{\tt arXiv:1602.08104}}].

\bibitem{Oriti:2016acw}
D.~Oriti, {\it {The universe as a quantum gravity condensate}},
  \href{http://arxiv.org/abs/1612.09521}{{\tt arXiv:1612.09521}}.

\bibitem{Oriti:2015rwa}
D.~Oriti, D.~Pranzetti, and L.~Sindoni, {\it {Horizon entropy from quantum
  gravity condensates}},  {\em Phys. Rev. Lett.} {\bf 116} (2016), no.~21
  211301, [\href{http://arxiv.org/abs/1510.06991}{{\tt arXiv:1510.06991}}].

\bibitem{Oriti:2016qtz}
D.~Oriti, L.~Sindoni, and E.~Wilson-Ewing, {\it {Emergent Friedmann dynamics
  with a quantum bounce from quantum gravity condensates}},  {\em Classical and
  Quantum Gravity} {\bf 33} (2016), no.~22 224001,
  [\href{http://arxiv.org/abs/1602.05881}{{\tt arXiv:1602.05881}}].

\bibitem{Rovelli:1990ph}
C.~Rovelli, {\it {What Is Observable in Classical and Quantum Gravity?}},  {\em
  Class. Quant. Grav.} {\bf 8} (1991) 297--316.

\bibitem{Brown:1994py}
J.~D. Brown and K.~V. Kucha\ifmmode~\check{r}\else \v{r}\fi{}, {\it {Dust as a
  standard of space and time in canonical quantum gravity}},  {\em Phys. Rev.}
  {\bf D51} (May, 1995) 5600--5629,
  [\href{http://arxiv.org/abs/gr-qc/9409001}{{\tt gr-qc/9409001}}].

\bibitem{Dittrich:2005kc}
B.~Dittrich, {\it {Partial and complete observables for canonical general
  relativity}},  {\em Class. Quant. Grav.} {\bf 23} (2006) 6155--6184,
  [\href{http://arxiv.org/abs/gr-qc/0507106}{{\tt gr-qc/0507106}}].

\bibitem{Ambjorn:1992us}
J.~Ambjorn, Z.~Burda, J.~Jurkiewicz, and C.~F. Kristjansen, {\it {3-d quantum
  gravity coupled to matter}},  {\em Phys. Lett.} {\bf B297} (1992) 253--260,
  [\href{http://arxiv.org/abs/hep-lat/9205021}{{\tt hep-lat/9205021}}].

\bibitem{Hamber:1993gn}
H.~W. Hamber and R.~M. Williams, {\it {Simplicial gravity coupled to scalar
  matter}},  {\em Nucl. Phys.} {\bf B415} (1994) 463--496,
  [\href{http://arxiv.org/abs/hep-th/9308099}{{\tt hep-th/9308099}}].

\bibitem{Ambjorn:1993ta}
J.~Ambjorn, Z.~Burda, J.~Jurkiewicz, and C.~F. Kristjansen, {\it {4-d quantum
  gravity coupled to matter}},  {\em Phys. Rev.} {\bf D48} (1993) 3695--3703,
  [\href{http://arxiv.org/abs/hep-th/9303042}{{\tt hep-th/9303042}}].

\bibitem{MoralesTecotl:1995jh}
H.~A. Morales-Tecotl and C.~Rovelli, {\it {Loop space representation of quantum
  fermions and gravity}},  {\em Nucl. Phys.} {\bf B451} (1995) 325--361.

\bibitem{Thiemann:1997rt}
T.~Thiemann, {\it {QSD 5: Quantum gravity as the natural regulator of matter
  quantum field theories}},  {\em Class. Quant. Grav.} {\bf 15} (1998)
  1281--1314, [\href{http://arxiv.org/abs/gr-qc/9705019}{{\tt gr-qc/9705019}}].

\bibitem{Oriti:2002bn}
D.~Oriti and H.~Pfeiffer, {\it {A Spin foam model for pure gauge theory coupled
  to quantum gravity}},  {\em Phys. Rev.} {\bf D66} (2002) 124010,
  [\href{http://arxiv.org/abs/gr-qc/0207041}{{\tt gr-qc/0207041}}].

\bibitem{Freidel:2004vi}
L.~Freidel and D.~Louapre, {\it {Ponzano-Regge model revisited I: Gauge fixing,
  observables and interacting spinning particles}},  {\em Class. Quant. Grav.}
  {\bf 21} (2004) 5685--5726, [\href{http://arxiv.org/abs/hep-th/0401076}{{\tt
  hep-th/0401076}}].

\bibitem{Freidel:2005cg}
L.~Freidel, D.~Oriti, and J.~Ryan, {\it {A Group field theory for 3-D quantum
  gravity coupled to a scalar field}},
  \href{http://arxiv.org/abs/gr-qc/0506067}{{\tt gr-qc/0506067}}.

\bibitem{Oriti:2006jk}
D.~Oriti and J.~Ryan, {\it {Group field theory formulation of 3-D quantum
  gravity coupled to matter fields}},  {\em Class. Quant. Grav.} {\bf 23}
  (2006) 6543--6576, [\href{http://arxiv.org/abs/gr-qc/0602010}{{\tt
  gr-qc/0602010}}].

\bibitem{Fairbairn:2006dn}
W.~J. Fairbairn, {\it {Fermions in three-dimensional spinfoam quantum
  gravity}},  {\em Gen. Rel. Grav.} {\bf 39} (2007) 427--476,
  [\href{http://arxiv.org/abs/gr-qc/0609040}{{\tt gr-qc/0609040}}].

\bibitem{Speziale:2007mt}
S.~Speziale, {\it {Coupling gauge theory to spinfoam 3d quantum gravity}},
  {\em Class. Quant. Grav.} {\bf 24} (2007) 5139--5160,
  [\href{http://arxiv.org/abs/0706.1534}{{\tt arXiv:0706.1534}}].

\bibitem{Dowdall:2010ej}
R.~J. Dowdall and W.~J. Fairbairn, {\it {Observables in 3d spinfoam quantum
  gravity with fermions}},  {\em Gen. Rel. Grav.} {\bf 43} (2011) 1263--1307,
  [\href{http://arxiv.org/abs/1003.1847}{{\tt arXiv:1003.1847}}].

\bibitem{Dowdall:2009ds}
R.~J. Dowdall, {\it {Wilson loops, geometric operators and fermions in 3d group
  field theory}},  {\em Central Eur. J. Phys.} {\bf 9} (2011) 1043--1056,
  [\href{http://arxiv.org/abs/0911.2391}{{\tt arXiv:0911.2391}}].

\bibitem{Bianchi:2010bn}
E.~Bianchi, M.~Han, C.~Rovelli, W.~Wieland, E.~Magliaro, and C.~Perini, {\it
  {Spinfoam fermions}},  {\em Class. Quant. Grav.} {\bf 30} (2013) 235023,
  [\href{http://arxiv.org/abs/1012.4719}{{\tt arXiv:1012.4719}}].

\bibitem{PhysRevD.93.024042}
J.~Lewandowski and H.~Sahlmann, {\it Loop quantum gravity coupled to a scalar
  field},  {\em Phys. Rev. D} {\bf 93} (Jan, 2016) 024042,
  [\href{http://arxiv.org/abs/1507.01149}{{\tt arXiv:1507.01149}}].

\bibitem{Fairbairn:2007sv}
W.~J. Fairbairn and E.~R. Livine, {\it {3d Spinfoam Quantum Gravity: Matter as
  a Phase of the Group Field Theory}},  {\em Class. Quant. Grav.} {\bf 24}
  (2007) 5277--5297, [\href{http://arxiv.org/abs/gr-qc/0702125}{{\tt
  gr-qc/0702125}}].

\bibitem{Girelli:2009yz}
F.~Girelli, E.~R. Livine, and D.~Oriti, {\it {4d Deformed Special Relativity
  from Group Field Theories}},  {\em Phys. Rev.} {\bf D81} (2010) 024015,
  [\href{http://arxiv.org/abs/0903.3475}{{\tt arXiv:0903.3475}}].

\bibitem{Oriti:2009pb}
D.~Oriti, {\it {Emergent non-commutative matter fields from Group Field Theory
  models of quantum spacetime}},  {\em J. Phys. Conf. Ser.} {\bf 174} (2009)
  012047, [\href{http://arxiv.org/abs/0903.3970}{{\tt arXiv:0903.3970}}].

\bibitem{Baratin:2010wi}
A.~Baratin and D.~Oriti, {\it {Group field theory with non-commutative metric
  variables}},  {\em Phys. Rev. Lett.} {\bf 105} (2010) 221302,
  [\href{http://arxiv.org/abs/1002.4723}{{\tt arXiv:1002.4723}}].

\bibitem{Baratin:2011hp}
A.~Baratin and D.~Oriti, {\it {Group field theory and simplicial gravity path
  integrals: A model for Holst-Plebanski gravity}},  {\em Phys. Rev.} {\bf D85}
  (Feb, 2012) 044003, [\href{http://arxiv.org/abs/1111.5842}{{\tt
  arXiv:1111.5842}}].

\bibitem{Guedes:2013vi}
C.~Guedes, D.~Oriti, and M.~Raasakka, {\it {Quantization maps, algebra
  representation and non-commutative Fourier transform for Lie groups}},  {\em
  J. Math. Phys.} {\bf 54} (2013) 083508,
  [\href{http://arxiv.org/abs/1301.7750}{{\tt arXiv:1301.7750}}].

\bibitem{Oriti:2014qoa}
D.~Oriti, {\it {Non-commutative quantum geometric data in group field
  theories}},  {\em Fortsch. Phys.} {\bf 62} (2014) 841--854,
  [\href{http://arxiv.org/abs/1405.1830}{{\tt arXiv:1405.1830}}].

\bibitem{Oriti:2014aka}
D.~Oriti and M.~Raasakka, {\it {Asymptotic Analysis of the Ponzano-Regge Model
  with Non-Commutative Metric Boundary Data}},  {\em SIGMA} {\bf 10} (2014)
  067, [\href{http://arxiv.org/abs/1401.5819}{{\tt arXiv:1401.5819}}].

\bibitem{Calcagni2013}
G.~Calcagni, D.~Oriti, and J.~Th\"{u}rigen, {\it Laplacians on discrete and
  quantum geometries},  {\em Classical and Quantum Gravity} {\bf 30} (2013),
  no.~12 125006, [\href{http://arxiv.org/abs/1208.0354}{{\tt
  arXiv:1208.0354}}].

\bibitem{Makela:2000ej}
J.~Makela and R.~M. Williams, {\it {Constraints on area variables in Regge
  calculus}},  {\em Class. Quant. Grav.} {\bf 18} (2001) L43,
  [\href{http://arxiv.org/abs/gr-qc/0011006}{{\tt gr-qc/0011006}}].

\bibitem{Calcagni:2012cv}
G.~Calcagni, D.~Oriti, and J.~Thurigen, {\it {Laplacians on discrete and
  quantum geometries}},  {\em Class. Quant. Grav.} {\bf 30} (2013) 125006,
  [\href{http://arxiv.org/abs/1208.0354}{{\tt arXiv:1208.0354}}].

\bibitem{Gielen2014a}
S.~Gielen, D.~Oriti, and L.~Sindoni, {\it Homogeneous cosmologies as group
  field theory condensates},  {\em Journal of High Energy Physics} {\bf 2014}
  (2014), no.~6 13, [\href{http://arxiv.org/abs/1311.1238}{{\tt
  arXiv:1311.1238}}].

\bibitem{PhysRevLett.111.031301}
S.~Gielen, D.~Oriti, and L.~Sindoni, {\it Cosmology from group field theory
  formalism for quantum gravity},  {\em Phys. Rev. Lett.} {\bf 111} (Jul, 2013)
  031301, [\href{http://arxiv.org/abs/1303.3576}{{\tt arXiv:1303.3576}}].

\bibitem{Oriti:2016ueo}
D.~Oriti, L.~Sindoni, and E.~Wilson-Ewing, {\it {Bouncing cosmologies from
  quantum gravity condensates}},  \href{http://arxiv.org/abs/1602.08271}{{\tt
  arXiv:1602.08271}}.

\bibitem{Cesare2017}
M.~de~Cesare and M.~Sakellariadou, {\it Accelerated expansion of the universe
  without an inflaton and resolution of the initial singularity from group
  field theory condensates},  {\em Physics Letters B} {\bf 764} (2017) 49 --
  53, [\href{http://arxiv.org/abs/1603.01764}{{\tt arXiv:1603.01764}}].

\bibitem{PhysRevD.94.064051}
M.~de~Cesare, A.~G.~A. Pithis, and M.~Sakellariadou, {\it Cosmological
  implications of interacting group field theory models: Cyclic universe and
  accelerated expansion},  {\em Phys. Rev. D} {\bf 94} (Sep, 2016) 064051,
  [\href{http://arxiv.org/abs/1606.00352}{{\tt arXiv:1606.00352}}].

\bibitem{PhysRevD.95.064004}
A.~G.~A. Pithis and M.~Sakellariadou, {\it Relational evolution of effectively
  interacting group field theory quantum gravity condensates},  {\em Phys. Rev.
  D} {\bf 95} (Mar, 2017) 064004, [\href{http://arxiv.org/abs/1612.02456}{{\tt
  arXiv:1612.02456}}].

\bibitem{Ashtekar2011}
A.~Ashtekar and P.~Singh, {\it Loop quantum cosmology: a status report},  {\em
  Classical and Quantum Gravity} {\bf 28} (2011), no.~21 213001,
  [\href{http://arxiv.org/abs/1108.0893}{{\tt arXiv:1108.0893}}].

\bibitem{Banerjee:2011qu}
K.~Banerjee, G.~Calcagni, and M.~Martin-Benito, {\it {Introduction to loop
  quantum cosmology}},  {\em SIGMA} {\bf 8} (2012) 016,
  [\href{http://arxiv.org/abs/1109.6801}{{\tt arXiv:1109.6801}}].

\bibitem{lrr-2008-4}
M.~Bojowald, {\it Loop quantum cosmology},  {\em Living Reviews in Relativity}
  {\bf 11} (2008), no.~4.

\bibitem{Gurau2011}
R.~Gurau, {\it Colored group field theory},  {\em Communications in
  Mathematical Physics} {\bf 304} (2011), no.~1 69--93,
  [\href{http://arxiv.org/abs/0907.2582}{{\tt arXiv:0907.2582}}].

\end{thebibliography}\endgroup

\appendix\label{sec:app}
\section{Order in $Z^{\Delta}_{\rm G}$ (Eq. (\ref{PureGravFeynExpan}))}\label{app:order}
In this appendix, we give an exact example of getting the right order for $P_{\rm G}$ and $V_{\rm G}$ in the Feynman amplitude (as in Eq. (\ref{PureGravFeynExpan})). Given a generic Feynman diagram $\Delta/\Delta_*$, the requried order can be obtained in two steps

\begin{enumerate}
\item One writes down all $V_{\rm G}$ associated with $\Delta$ in a row, to make sure for every face, the order of the associated vertices (read by following the boundary loop of the face) coincides with the order of the same vertices in the row, either from left to right or the opposite, given the first $V_{\rm G}$ is considered to be connected to the right of the last $V_{\rm G}$ in the row.
\item One after one from the left to the right in the row, for every $V_{\rm G}$, one connects all the $P_{\rm G}$ that touch that $V_{\rm G}$  according to $\Delta$ with the $\star$-products after that $V_{\rm G}$, in an order that, given $P_{\rm G}\backslash P_{\rm G}'$ connected to the vertex function $V_{\rm G}$ and the other vertex functions they connect to being denoted as $V_{\rm G}^{\rm other} \backslash {V_{\rm G}^{\rm other}}'$, then if $V_{\rm G}^{\rm other}$ is to the right of ${V_{\rm G}^{\rm other}}'$ in the row, $P_{\rm G}$ is to be put in the left of $P_{\rm G}'$
    ; and if several $P_{\rm G}$ all connect to the same $V_{\rm G}^{\rm other}$, the order is free to choose. The point is to make sure given a face, the $P_{\rm G}$ within this face are properly lined up according to the boundary links of that face.
\end{enumerate}

\begin{figure}
\centering
\includegraphics{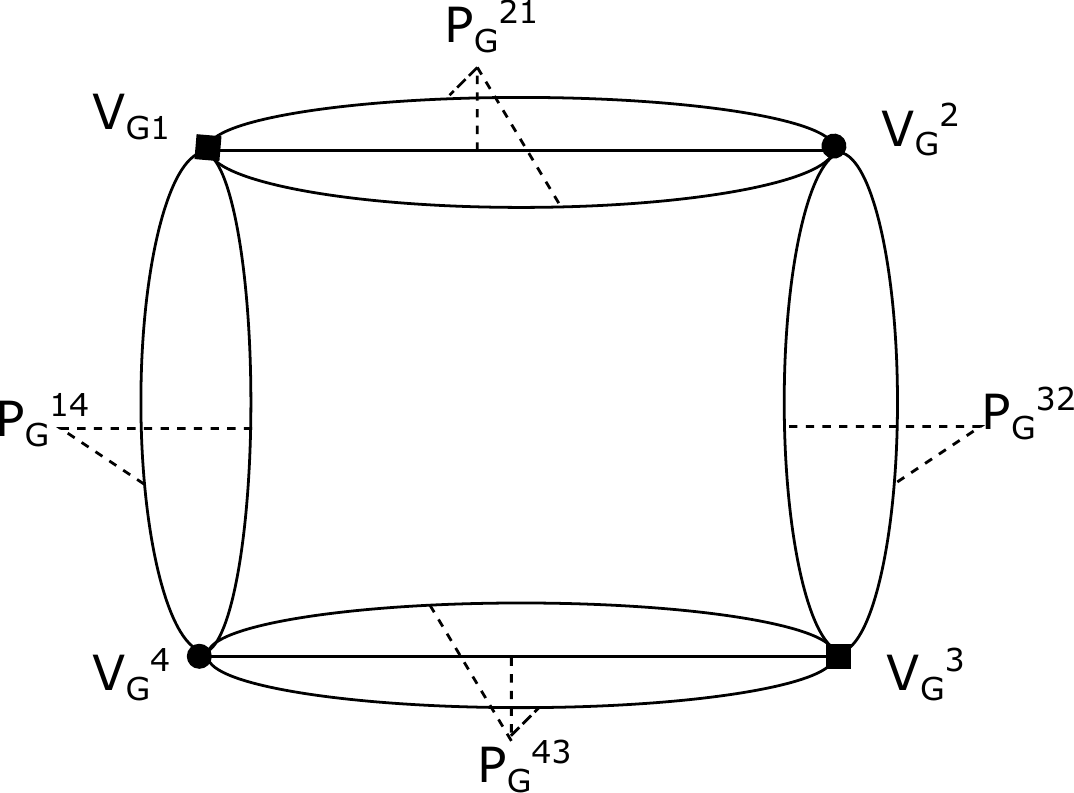}
\caption{The Feynman diagram ($2$-complex) with $4$ vertices and $8$ links.}
\label{FeynDia}
\end{figure}

Then for a diagram as shown in Fig. \ref{FeynDia}, we follow these steps. So firstly, we write down the row of the four $V_{\rm G}$ under the convolution:
\begin{eqnarray}
Z^{\Delta}_{\rm G}= \int_{\Delta}{V}_{\rm G}^4\star\ldots\star V_{\rm G}^3 \star\ldots\star {V}_{\rm G}^2 \star\ldots\star V_{\rm G}^1,
\end{eqnarray}
where we do not distinguish between the two different vertices in the diagram since ${V}_{\rm G}=\bar{V}_{\rm G}$ as we choose. Then we need to fill in the propagator $P_{\rm G}$ for every ${V}_{\rm G}$ from the left to the right.\\
\indent For $V_{\rm G}^4$, we filled in as:
\begin{eqnarray}
Z^{\Delta}_{\rm G}=\int_{\Delta}{V}_{\rm G}^4\star P_{\rm G}^{14}\star P_{\rm G}^{14}\star P_{\rm G}^{43}\star P_{\rm G}^{43}\star P_{\rm G}^{43}\star V_{\rm G}^3 \star\ldots\star {V}_{\rm G}^2 \star\ldots\star V_{\rm G}^1.
\end{eqnarray}
\indent For $V_{\rm G}^3$, we fill in as:
\begin{eqnarray}
Z^{\Delta}_{\rm G}=\int_{\Delta}{V}_{\rm G}^4\star P_{\rm G}^{14}\star P_{\rm G}^{14}\star P_{\rm G}^{43}\star P_{\rm G}^{43}\star P_{\rm G}^{43}\star V_{\rm G}^3 \star P_{\rm G}^{32}\star P_{\rm G}^{32}\star {V}_{\rm G}^2 \star\ldots\star V_{\rm G}^1.
\end{eqnarray}
\indent For $V_{\rm G}^3$, fill in as:
\begin{eqnarray}
Z^{\Delta}_{\rm G}=\int_{\Delta} {V}_{\rm G}^4\star P_{\rm G}^{14}\star P_{\rm G}^{14}\star P_{\rm G}^{43}\star P_{\rm G}^{43}\star P_{\rm G}^{43}\star V_{\rm G}^3 \star P_{\rm G}^{32}\star P_{\rm G}^{32}\star {V}_{\rm G}^2 \star P_{\rm G}^{21} \star P_{\rm G}^{21}\star P_{\rm G}^{21}\star V_{\rm G}^1.
\end{eqnarray}
\indent Here the upper indexes of $P_{\rm G}$ denote the two vertices it connects. Note we do not explicitly show the (convoluted) arguments of $P_{\rm G}$ and $V_{\rm G}$ or the faces in Fig. \ref{FeynDia}, both of which one needs to check the specific gluing of the triangles of the dual simplcial complex to be able to tell, but the point is for a space-time (pseudo-)manifold triangulation, following the boundary loop of a face of the dual $2$-complex, every given link (and so the $P_{\rm G}$ associated with that link) only comes up once, wherefore our order in Eq. (\ref{PureGravFeynExpan}) makes sure the face holonomy to be formed correctly for every face of the $2$-complex that rebuilds a space-time (pseudo-)manifold.\\
\indent Following the standard Feynman expansion in GFT, this diagram in Fig. \ref{FeynDia} is produced from
\begin{eqnarray}
\hat{\bar{V}}_{\rm G}^4\hat{{V}}_{\rm G}^3\hat{\bar{V}}_{\rm G}^2\hat{{V}}_{\rm G}^1e^{\int [dx'^4][dx^4]{\rm J}(\vec{x}')\star P_{\rm G}(\vec{x}',\vec{x})\star{\rm \bar{J}}(\vec{x})},
\end{eqnarray}
where
\begin{eqnarray}
\hat{\bar{V}}_{\rm G}^i&=&{V}_{\rm G}(\vec{x}^i_1, \vec{x}^i_2,\vec{x}^i_3,\vec{x}^i_4,\vec{x}^i_5)\star\frac{\delta}{\delta {\rm \bar{J}}(\vec{x}^i_1)}\star\frac{\delta}{\delta {\rm \bar{J}}(\vec{x}^i_2)}\star\frac{\delta}{\delta {\rm \bar{J}}(\vec{x}^i_3)}\star\frac{\delta}{\delta {\rm \bar{J}}(\vec{x}^i_4)}\star\frac{\delta}{\delta {\rm \bar{J}}(\vec{x}^i_5)}, i=2,4\\
\hat{{V}}_{\rm G}^i&=&{V}_{\rm G}(\vec{x}^i_1, \vec{x}^i_2,\vec{x}^i_3,\vec{x}^i_4,\vec{x}^i_5)\star\frac{\delta}{\delta {\rm {J}}(\vec{x}^i_1)}\star\frac{\delta}{\delta {\rm {J}}(\vec{x}^i_2)}\star\frac{\delta}{\delta {\rm {J}}(\vec{x}^i_3)}\star\frac{\delta}{\delta {\rm {J}}(\vec{x}^i_4)}\star\frac{\delta}{\delta {\rm {J}}(\vec{x}^i_5)}, i=1,3
\end{eqnarray}
and ${\rm J}$ is the source field. \\
\indent However the GFT Feynamn expansion produces more than just the triangulation of the space-time (pseudo-)manifold: as the GFT Feynman diagram actually corresponds to the arbitrary gluing of the simplices, some of the gluing (as simplicial complex) do not correspond to a (pseudo-)manifold due to the topological singularities \cite{Gurau2011}.

\section{Geometrical functions $\tilde{V}_l$, $L_l$ and $V_v$}\label{app:VLV}
In this appendix, we derive the simplification for $\tilde{V}_l$, $L_l$ and $V_v$ as functions of $x$ under the continuum limit. The goal is to simplified these functions to depend only on a single tetrahedron (at least for $\tilde{V}_l$ and $L_l$), and the strategy is to firstly derive the (exact) $3$-volume of this tetrahedron, which we then use to derive the simplified $\tilde{V}_l$, $L_l$ and $V_v$ (as $\tilde{V}$, $L$ and $V$) under the continuum limit.\\
\indent Firstly we define $\tilde{x}$:
\begin{eqnarray}
\tilde{x}=\frac{1}{\gamma^2 - 1}(\gamma^2 x- \gamma (*x)),
\end{eqnarray}
where $(*x)^{QI}=\frac{1}{4}\epsilon_{QIJK}x^{JK}$ is the conjugate of $x$ as a bivector. Then $\tilde{x}$ is related to the discrete geometry as:
\begin{eqnarray}
\tilde{x}_i^{IJ}=N^I\wedge n_i^J,
\end{eqnarray}
where $i$ denotes four triangles of a tetrahedron and the capital index is the vector index. Thus $n_i^J$ is the $4$-vector orthogonal to the triangle $i$, and $N^I$ is the normalized $4$-vector orthogonal to the whole tetrahedron:
\begin{eqnarray}
N_I N^I&=&1\\
N_I n_i^I&=&0,
\end{eqnarray}
where by default, the repeated capital case index $I$ is summed from $1$ to $4$.\\
\indent We also have
\begin{eqnarray}
n_{i}^I&=&\sum_{i_0,i_1,i_2}\frac{1}{2\times 3!}\epsilon_i^{\,\,i_0i_1i_2}\epsilon^I_{\,\,I_1I_2}e_{i_0i_1}^{I_1}e_{i_0i_2}^{I_2},
\end{eqnarray}
where $e_{i_0i_1}^{I_1}$ is the vector of the edge shared by triangle $i_0$ and $i_1$ of the tetrahedron. \\
\indent Then we construct
\begin{eqnarray}
X&=&\sum_{i,j,k,q}\frac{1}{2}\epsilon^{ijkq}\epsilon_{QIJK}\tilde{x}_{v_i}^{QI}\tilde{x}_{v_j}^{JM}\tilde{x}_{{v_k}M}^{\,\,\,\,\,\,\,\,\,\,\,\,K}\nonumber \\
&=& \sum_{i,j,k,q}\epsilon^{ijkq}\epsilon_{IJK}n_i^{I}n_j^Jn_k^{\,\,K}\nonumber\\
&=&(\frac{1}{2\times 3!})^3(T_1+T_2+T_3+T_4+T_5+T_6),
\end{eqnarray}
where
\begin{eqnarray}
T_1&=&\sum_{(\ldots)}(\delta ^{q i_0} \epsilon ^{i_1 j_0 j_1 j_2} \epsilon ^{i_2 k_0 k_1 k_2} \epsilon _{I_1 J_1 J_2} \epsilon _{I_2 K_1 K_2}) E^{I_1I_2J_1J_2K_1K_2}_{i_0i_1i_2j_0j_1j_2k_0k_1k_2}-\nonumber \\
&&\sum_{(\ldots)}(\delta ^{q i_0} \epsilon ^{i_2 j_0 j_1 j_2} \epsilon ^{i_1 k_0 k_1 k_2} \epsilon _{I_1 J_1 J_2} \epsilon _{I_2 K_1 K_2}) E^{I_1I_2J_1J_2K_1K_2}_{i_0i_1i_2j_0j_1j_2k_0k_1k_2}\\
T_2&=&\sum_{(\ldots)}(\delta ^{q i_0} \epsilon ^{i_2 j_0 j_1 j_2} \epsilon ^{i_1 k_0 k_1 k_2} \epsilon _{I_2 J_1 J_2} \epsilon _{I_1 K_1 K_2}) E^{I_1I_2J_1J_2K_1K_2}_{i_0i_1i_2j_0j_1j_2k_0k_1k_2}-\nonumber\\
&&\sum_{(\ldots)}(\delta ^{q i_0} \epsilon ^{i_1 j_0 j_1 j_2} \epsilon ^{i_2 k_0 k_1 k_2} \epsilon _{I_2 J_1 J_2} \epsilon _{I_1 K_1 K_2}) E^{I_1I_2J_1J_2K_1K_2}_{i_0i_1i_2j_0j_1j_2k_0k_1k_2}\\
T_3&=&\sum_{(\ldots)}(\delta ^{q i_1} \epsilon ^{i_2 j_0 j_1 j_2} \epsilon ^{i_0 k_0 k_1 k_2} \epsilon _{I_1 J_1 J_2} \epsilon _{I_2 K_1 K_2}) E^{I_1I_2J_1J_2K_1K_2}_{i_0i_1i_2j_0j_1j_2k_0k_1k_2}-\nonumber \\
&&\sum_{(\ldots)}(\delta ^{q i_2} \epsilon ^{i_1 j_0 j_1 j_2} \epsilon ^{i_0 k_0 k_1 k_2} \epsilon _{I_1 J_1 J_2} \epsilon _{I_2 K_1 K_2}) E^{I_1I_2J_1J_2K_1K_2}_{i_0i_1i_2j_0j_1j_2k_0k_1k_2}\\
T_4&=&\sum_{(\ldots)}(\delta ^{q i_2} \epsilon ^{i_1 j_0 j_1 j_2} \epsilon ^{i_0 k_0 k_1 k_2} \epsilon _{I_2 J_1 J_1} \epsilon _{I_1 K_1 K_2}) E^{I_1I_2J_1J_2K_1K_2}_{i_0i_1i_2j_0j_1j_2k_0k_1k_2}-\nonumber \\
&&\sum_{(\ldots)}(\delta ^{q i_1} \epsilon ^{i_2 j_0 j_1 j_2} \epsilon ^{i_0 k_0 k_1 k_2} \epsilon _{I_2 J_1 J_2} \epsilon _{I_1 K_1 K_2}) E^{I_1I_2J_1J_2K_1K_2}_{i_0i_1i_2j_0j_1j_2k_0k_1k_2}\\
T_5&=&\sum_{(\ldots)}(\delta ^{q i_2} \epsilon ^{i_0 j_0 j_1 j_2} \epsilon ^{i_1 k_0 k_1 k_2} \epsilon _{I_1 J_1 J_2} \epsilon _{I_2 K_1 K_2}) E^{I_1I_2J_1J_2K_1K_2}_{i_0i_1i_2j_0j_1j_2k_0k_1k_2}-\nonumber \\
&&\sum_{(\ldots)}(\delta ^{q i_1} \epsilon ^{i_0 j_0 j_1 j_2} \epsilon ^{i_2 k_0 k_1 k_2} \epsilon _{I_1 J_1 J_2} \epsilon _{I_2 K_1 K_2}) E^{I_1I_2J_1J_2K_1K_2}_{i_0i_1i_2j_0j_1j_2k_0k_1k_2}\\
T_6&=&\sum_{(\ldots)}(\delta ^{q i_1} \epsilon ^{i_0 j_0 j_1 j_2} \epsilon ^{i_2 k_0 k_1 k_2} \epsilon _{I_2 J_1 J_2} \epsilon _{I_1 K_1 K_2}) E^{I_1I_2J_1J_2K_1K_2}_{i_0i_1i_2j_0j_1j_2k_0k_1k_2}-\nonumber \\
&&\sum_{(\ldots)}(\delta ^{q i_2} \epsilon ^{i_0 j_0 j_1 j_2} \epsilon ^{i_1 k_0 k_1 k_2} \epsilon _{I_2 J_1 J_2} \epsilon _{I_1 K_1 K_2}) E^{I_1I_2J_1J_2K_1K_2}_{i_0i_1i_2j_0j_1j_2k_0k_1k_2}
\end{eqnarray}
and
\begin{eqnarray}
E^{I_1I_2J_1J_2K_1K_2}_{i_0i_1i_2j_0j_1j_2k_0k_1k_2}\equiv e^{I_1}_{i_0 i_1} e^{I_2}_{i_0 i_2} e^{J_1}_{j_0 j_1} e^{J_2}_{j_0 j_2} e^{K_1}_{k_0 k_1} e^{K_2}_{k_0 k_2}.
\end{eqnarray}
Here $(\ldots)\equiv(q, i, i_0, i_1, i_2, j, j_0, j_1, j_2, k, k_0, k_1, k_2,)$ denotes the indices needed to be summed (from $1$ to $4$). Besides the repeated capital case indices $I_1, I_2, J_1, J_2, K_1, K_2$ are also summed from $1$ to $4$ by default. So $X$ only depends on a tetrahedron.\\
\indent Then by carefully evaluating $X$, we can find
\begin{eqnarray}
|V_{(3)}|=\frac{1}{6\sqrt{2}}|X^{\frac{1}{2}}|,
\end{eqnarray}
where $V_{(3)}$ is the $3$-volume of the tetrahedron, with the help of
\begin{eqnarray}
|V_{(3)}|=\frac{1}{6\times4!}|\sum_{i,j,k,q}\epsilon^{qijk}\epsilon_{IJK}e_{qi}^I e_{lj}^J e_{lk}^K|.
\end{eqnarray}
\indent Now bearing in mind that the continuum limit means the equilateral limit of the lattice, then from
\begin{eqnarray}
|V_{(n)}|=\frac{a^n}{n!}\sqrt{\frac{n+1}{2^n}},
\end{eqnarray}
where $V_{(n)}$ is the volume of $n$-simplex with the equal edge length $a$, for $\tilde{V}_l$ and $V_v$'s simplification $\tilde{V}$ and $V$, we have
\begin{eqnarray}
|V|&=&|V_{(4)}|=|\frac{\sqrt{5}\sqrt[3]{3}}{8}V_{(3)}^{\frac{4}{3}}|\\
|\tilde{V}|&=&\frac{2}{5}|V_{(4)}|=|\frac{\sqrt[3]{3}}{4\sqrt{5}}V_{(3)}^{\frac{4}{3}}|
\end{eqnarray}
\indent Also in this limit, we can take $L_l$'s simplification $L$ as two times of the radius of the inscribed sphere to the 4-simplex:
\begin{eqnarray}
L=2\left|\frac{4V_{(4)}}{5V_{(3)}}\right|
\end{eqnarray}
\indent So finally, we have
\begin{eqnarray}
\tilde{V}&=&{\rm Sgn}\left(X\right)\frac{1}{48\sqrt{5}}|X^{\frac{2}{3}}|\\
L&=&\frac{1}{\sqrt{10}}|X^{\frac{1}{6}}|\\
V&=&{\rm Sgn}\left(X\right)\frac{\sqrt{5}}{96}|X^{\frac{2}{3}}|,
\end{eqnarray}
where we use ${\rm Sgn}\left(X\right)$ to indicate the orientation of $\tilde{V}$ and $V$.
\section{$\star$-product correction}\label{app:correction}
In this appendix, we calculate the correction of the $\star$-product to the normal product for general functions. \\
\indent Firstly, let us assume we have ${\rm O}_e(\hbar)$ as the correction of the $\star$-product between the plane waves:
\begin{eqnarray}
\left(e_{g}\star e_{g'}\right)(x) =e_{g}(x) e_{g'}(x)\left(1 + {\rm O}_e(\hbar)\right).
\end{eqnarray}
\indent Then by using the group Fourier transform, we can derive the correction ${\rm O}(\hbar)$ for the general functions ${\rm f}$ and ${\rm f}'$:
\begin{eqnarray}
\left({\rm f}\star {\rm f}'\right)(x)&=&\int dg {\rm f}(g)e_{g}(x)\star\int dg' {\rm f}'(g')e_{g'}(x)\nonumber\\
&=&\int dg {\rm f}(g)\int dg' {\rm f}'(g')\left(e_{g}\star e_{g'}\right)(x)\nonumber\\
&=&\int dg {\rm f}(g)e_{g}(x)\int dg' {\rm f}'(g') e_{g'}(x)\left(1 + {\rm O}_e(\hbar)\right)\nonumber\\
&=&\int dg {\rm f}(g)e_{g}(x)\int dg' {\rm f}'(g') e_{g'}(x)+\int dg dg'{\rm f}(g)e_{g}(x) {\rm f}'(g')e_{g'}(x){\rm O}_e(\hbar)\nonumber \\
&=&{\rm f}(x){\rm f}'(x)(1+{\rm O}(\hbar)),
\end{eqnarray}
where
\begin{eqnarray}
{\rm O}(\hbar)=\frac{1}{{\rm f}(x){\rm f}'(x)}\int dg dg'{\rm f}(g)e_{g}(x) {\rm f}'(g')e_{g'}(x){\rm O}_e(\hbar).
\end{eqnarray}
\indent So in order to find ${\rm O}(\hbar)$, now we only have to find ${\rm O}_e(\hbar)$. In principle, given the definition of the plane wave one can always write
\begin{eqnarray}
{\rm O}_e(\hbar)=\frac{\left(e_{g}\star e_{g'}\right)(x)}{e_{g}(x) e_{g'}(x)}-1,
\end{eqnarray}
and then expand it to any desired order of $\hbar$ as needed.\\
\indent Here we will be more concrete by taking group $G={\rm SU}(2)$ as an example. So the group element $g$ can be parameterized as
\begin{eqnarray}
g=e^{\tilde{\kappa}\vec{k}\cdot\vec{\tau}},
\end{eqnarray}
where $\vec{\tau}=i\vec{\sigma}$, $\sigma_1$, $\sigma_2$, $\sigma_3$ are the three Pauli matrices; and $\tilde{\kappa}=\hbar\kappa$, $\kappa$ is the bare Newton constant. And so the ${\mathfrak s}{\mathfrak u}(2)$ element is written as $x=\vec{x}\cdot \vec{\tau}$.\\
\indent For the \textit{Freidel}-\textit{Livine}-\textit{Majid} case, the definition of the plane wave is
\begin{eqnarray}
e^{\rm FLM}_g(x)=e^{\frac{i}{2 \tilde{\kappa}}{\rm Tr}(xg)}.
\end{eqnarray}
\indent So we have
\begin{eqnarray}
{\rm Tr}(xg)=2 \sin(\tilde{\kappa}|\vec{k}|)\vec{n}_k\cdot\vec{x},
\end{eqnarray}
where $|\vec{k}|$ is the module of $\vec{k}$ and $\vec{n}_k=\frac{\vec{k}}{|\vec{k}|}$.\\
\indent Then one finds
\begin{eqnarray}
{\rm Tr}(xgg')={\rm Tr}(xg)+{\rm Tr}(xg')+\Omega,
\end{eqnarray}
where
\begin{eqnarray}
\Omega&=&\left(\sum_{n=1}\frac{(\tilde{\kappa}|\vec{k}|)^{2n}}{(2n)!}\right)\sin({\tilde{\kappa} |\vec{k'}|})\vec{n}_k'\cdot\vec{x}+\left(\sum_{n=1}\frac{(\tilde{\kappa}|\vec{k'}|)^{2n}}{(2n)!}\right)\sin({\tilde{\kappa} |\vec{k}|})\vec{n}_k\cdot\vec{x}+\nonumber\\
&&\sin({\tilde{\kappa} |\vec{k}|})\sin({\tilde{\kappa}|\vec{k}'|})\sum_{a, b, c=1}^{3}x^an_k^b{n'}_k^c\epsilon_{abc}.
\end{eqnarray}
\indent So finally for $e^{\rm FLM}_g(x)$, we have
\begin{eqnarray}
\left(e^{\rm FLM}_g\star e^{\rm FLM}_{g'}\right)(x)&=&e^{\rm FLM}_{gg'}(x)\nonumber\\
&=&e^{\frac{i}{2 \tilde{\kappa}}{\rm Tr}(xgg')}\nonumber\\
&=&e^{\frac{i}{2 \tilde{\kappa}}\left({\rm Tr}(xg)+{\rm Tr}(xg')+\Omega\right)}\nonumber\\
&=&e^{\frac{i}{2 \tilde{\kappa}}{\rm Tr}(xg)}e^{\frac{i}{2 \tilde{\kappa}}{\rm Tr}(xg')}e^{\frac{i}{2 \tilde{\kappa}}\Omega}\nonumber\\
&=&e^{\rm FLM}_g(x) e^{\rm FLM}_{g'}(x)\left(1+{\rm O}^{\rm FLM}_e(\hbar)\right),
\end{eqnarray}
where
\begin{eqnarray}
{\rm O}^{\rm FLM}_e(\hbar)=\sum_{n=1}\frac{1}{n!}\left(\frac{i}{2\tilde{\kappa}}\Omega\right)^n.
\end{eqnarray}
\indent For the \textit{Duflo} case, plane wave is defined as
\begin{eqnarray}
e^{\rm D}_g(x)=\frac{\sin(\tilde{\kappa}|\vec{k}|)}{\tilde{\kappa}|\vec{k}|}e^{\frac{i}{2 \tilde{\kappa}}(\tilde{\kappa}\vec{k}\cdot\vec{x})}.
\end{eqnarray}
\indent Then supposing
\begin{eqnarray}
gg'=e^{\tilde{\kappa}\vec{k}\cdot\vec{\tau}}e^{\tilde{\kappa}\vec{k}'\cdot\vec{\tau}}=e^{\tilde{\kappa}\vec{k}_{gg'}\cdot\vec{\tau}},
\end{eqnarray}
we find
\begin{eqnarray}
\vec{k}_{gg'}=\vec{k}+\vec{k}'-2\tilde{\kappa}\vec{k}\times\vec{k}'.
\end{eqnarray}
\indent So we have
\begin{eqnarray}
\left(e^{\rm D}_g\star e^{\rm D}_{g'}\right)(x)&=&e^{\rm D}_{gg'}(x)\nonumber\\
&=&\frac{\sin(\tilde{\kappa}|\vec{k}_{gg'}|)}{\tilde{\kappa}|\vec{k}_{gg'}|}e^{\frac{i}{2 \tilde{\kappa}}(\tilde{\kappa}\vec{k}_{gg'}\cdot\vec{x})}\nonumber\\
&=&\frac{\sin(\tilde{\kappa}|\vec{k}_{gg'}|)}{\tilde{\kappa}|\vec{k}_{gg'}|}e^{\frac{i}{2 \tilde{\kappa}}(\tilde{\kappa}\left(\vec{k}+\vec{k}'-2\tilde{\kappa}\vec{k}\times\vec{k}'\right)\cdot\vec{x})}\nonumber\\
&=&\frac{\sin(\tilde{\kappa}|\vec{k}_{gg'}|)}{\tilde{\kappa}|\vec{k}_{gg'}|}e^{\frac{i}{2 \tilde{\kappa}}(\tilde{\kappa}\vec{k}\cdot\vec{x})}e^{\frac{i}{2 \tilde{\kappa}}(\tilde{\kappa}\vec{k}'\cdot\vec{x})}e^{-i\tilde{\kappa}\vec{k}\times\vec{k}'\cdot\vec{x}}\nonumber\\
&=&e^{\rm D}_g(x) e^{\rm D}_{g'}(x)\Theta\left(1+{{\rm O}}^D_1(\hbar)\right),
\end{eqnarray}
where
\begin{eqnarray}
\Theta&=&\frac{\tilde{\kappa}\sin(\tilde{\kappa}|\vec{k}_{gg'}|)|\vec{k}||\vec{k}'|}{\sin(\tilde{\kappa}|\vec{k}|)\sin(\tilde{\kappa}|\vec{k}'|)|\vec{k}_{gg'}|}\nonumber\\
{{\rm O}}^D_1(\hbar)&=&\sum_{n=1}\frac{1}{n!}\left(-i\tilde{\kappa}(\vec{k}\times\vec{k}')\cdot\vec{x}\right)^n.
\end{eqnarray}
\indent Then by expanding $\Theta$ with respect to $\hbar$, one finds $\Theta$ can be written as
\begin{eqnarray}
\Theta=1+{\rm O}^D_2(\hbar^2),
\end{eqnarray}
wherefore
\begin{eqnarray}
\left(e^{\rm D}_g\star e^{\rm D}_{g'}\right)(x)&=&e^{\rm D}_g(x) e^{\rm D}_{g'}(x)\Theta\left(1+{{\rm O}}^D_1(\hbar)\right)\nonumber\\
&=&e^{\rm D}_g(x) e^{\rm D}_{g'}(x)\left(1+{\rm O}^D_2(\hbar^2)\right)\left(1+{{\rm O}}^D_1(\hbar)\right)\nonumber\\
&=&e^{\rm D}_g(x) e^{\rm D}_{g'}(x)\left(1+{{\rm O}}^D_1(\hbar)+{{\rm O}}^D_2(\hbar^2)+{{\rm O}}^D_1(\hbar){{\rm O}}^D_2(\hbar^2)\right),
\end{eqnarray}
where
\begin{eqnarray}
{\rm O}^D_2(\hbar^2)=\Theta-1=\frac{\tilde{\kappa}\sin(\tilde{\kappa}|\vec{k}_{gg'}|)|\vec{k}||\vec{k}'|}{\sin(\tilde{\kappa}|\vec{k}|)\sin(\tilde{\kappa}|\vec{k}'|)|\vec{k}_{gg'}|}-1.
\end{eqnarray}
\indent So finally for $e^{\rm D}_g(x)$, we have
\begin{eqnarray}
{{\rm O}}^D_e(\hbar)={{\rm O}}^D_1(\hbar)+{{\rm O}}^D_2(\hbar^2)+{{\rm O}}^D_1(\hbar){{\rm O}}^D_2(\hbar^2).
\end{eqnarray}
\section{$2N$th order truncation}\label{app:truncation}
In this appendix, we look into the some general mathematical discussion of the $\tilde{\mathcal P}_{{\phi }(2N)}^{-1}$ and ${\mathcal P}_{{\phi }(2N)}$.\\
\indent For
\begin{eqnarray}
\tilde{\mathcal P}_{{\phi }(2N)}^{-1}=\sum_{n=0}^{N}b_{2n}\partial_{\phi}^{2n},
\end{eqnarray}
we compute ${\mathcal P}_{{\phi }(2N)}$ as
\begin{eqnarray}\label{genppphi}
{\mathcal P}_{{\phi }(2N)}&=&\int dp \frac{e^{-i p (\phi-\phi')}}{b_{2N}\prod_{i=1}^{\tilde{N}}(p^2-r_i^2)^{N_i}},
\end{eqnarray}
where $p=\pm r_i$ are the roots of equation $\sum_{n=0}^{N}(-1)^nb_{2n}p^{2n}=0$; and $\sum_{i=1}^{\tilde{N}}N_i=N$, $N>1$ for we have discussed $N=1$ case. \\
\indent As in the previous discussion, we will only consider $r_i\neq0$ since we want our theory to be able to return to the pure gravity GFT; and we consider ${\rm Im}(r_i)\neq0$ since the condition ${\rm Im}(r_i)=0$ can not hold valid when we have to integrate both orientations of the volume element in the Feynaman amplitude, which can as well be seen from the previously discussed $N=2$ case as an example.\\
\indent Then we can derive ${\mathcal P}_{{\phi }(2N)}$ using the contour integral. Firstly, we define $R_i={\rm SI}(r_i)r_i$. Then starting with the simple case, where all $N_i=1$ and $\tilde{N}=N$ (which equally says all $r_i$ are the first order roots), we have,
\begin{eqnarray}
{\mathcal P}_{{\phi }(2N)}&=&\frac{-2i\pi }{b_{2N}}\sum_{i=1}^{N}(\frac{e^{-i|\phi-\phi'| R_i}}{2R_i\prod_{j\neq i}^N(R_i^2-R_j^2)})\nonumber\\
&=&\frac{-2i\pi }{b_{2N}(\prod_{1\leqslant i<j\leqslant N}(R_i^2-R_j^2))}\left(\sum_{i=1}^{N}\frac{(-1)^{i-1}e^{-i|\phi-\phi'| R_i}\prod_{1\leqslant q(\neq i)<j(\neq i)\leqslant N}(R_q^2-R_j^2)}{2R_i}\right)\nonumber\\
&=&\sum_{\tilde{m}} \tilde{c}_{\tilde{m}}|\phi-\phi'|^{\tilde{m}},
\end{eqnarray}
where the expansion of $|\phi-\phi'|$ is of the same meaning as in the previous discussion, and
\begin{eqnarray}
\tilde{c}_{\tilde{m}}&=&\frac{-2i\pi }{b_{2N}(\prod_{1\leqslant i<j\leqslant N}(R_i^2-R_j^2))}(\sum_{i=1}^{N}\frac{(-1)^{i-1}\frac{(-i R_i)^{\tilde{m}}}{\tilde{m}!}\prod_{1\leqslant q(\neq i)<j(\neq i)\leqslant N}(R_q^2-R_j^2)}{2R_i})\nonumber \\
&=&\frac{(-i)^{\tilde{m}+1}\pi }{b_{2N} \tilde{m}!(\prod_{1\leqslant i<j\leqslant N}(R_i^2-R_j^2))}{\det}\left(
                                                                                   \begin{array}{cccccc}
                                                                                     R_1^{\tilde{m}-1} & R_2^{\tilde{m}-1}  & \cdots & R_N^{\tilde{m}-1} \\
                                                                                     R_1^{2N-4} & R_2^{2N-4}  & \cdots & R_N^{2N-4} \\
                                                                                     \vdots & \vdots  & \ddots & \vdots \\
                                                                                     R_1^4 & R_2^4  & \cdots & R_N^4 \\
                                                                                     R_1^2 & R_2^2  & \cdots & R_N^2 \\
                                                                                     1 & 1  & \cdots & 1 \\
                                                                                   \end{array}
                                                                                 \right)_{N \times N},
\end{eqnarray}
where we can see $\tilde{c}_{\tilde{m}}=0$ when $\tilde{m}=1, 3, ..., (2N-3)$, and $\tilde{c}_{\tilde{m}}\neq0$ when $\tilde{m}$ is even.\\
\indent As for the case where not all $r_i$ are first order roots, for the coefficients of the expansion ${\mathcal P}_{{\phi }(2N)}=\sum_{\tilde{m}} \tilde{c}_{\tilde{m}}|\phi-\phi'|^{\tilde{m}}$, we have a conjecture:
\begin{eqnarray}
\tilde{c}_{\tilde{m}}=\frac{(-i)^{\tilde{m}+1}\pi}{\rho  \tilde{m}!(\prod_i^{\tilde{N}} R_i^{N_i(N_i-1)})(\prod_{1\leqslant i<j}(R_i^2-R_j^2)^{N_iN_j})}{\det}\left(
                              \begin{array}{c}
                                u(\tilde{m}-1) \\
                                u(2N-4)\\
                                \vdots \\
                                u(2) \\
                                u(0) \\
                              \end{array}
                            \right)_{N \times N},
\end{eqnarray}
where $\rho$ is a constant depending on the case (meaning $N_i$ and $\tilde{N}$), and
\begin{eqnarray}
u(t)=\left(u_1(t), u_2(t), \cdots, u_{\tilde{N}}(t)\right).
\end{eqnarray}
When $N_i>1$,
\begin{eqnarray}
u_i(t)=\left(
       \begin{array}{ccccc}
          R_i^t, & (\frac{2-t}{2})^{N_i-1}R_i^t, & (\frac{2-t}{2})^{N_i-2}(\frac{4-t}{4})R_i^t, & ..., & \prod_{j=1}^{N_i-1}\frac{2j-t}{2j}R_i^t \\
       \end{array}
     \right);
\end{eqnarray}
When $N_i=1$,
\begin{eqnarray}
u_i(t)=\left(
       \begin{array}{c}
          R_{i}^t  \\
       \end{array}
     \right).
\end{eqnarray}
\indent This conjecture has been tested to be true for several simple cases. So still, we see $\tilde{c}_{\tilde{m}}=0$ when $\tilde{m}=1, 3, ..., (2N-3)$, and $\tilde{c}_{\tilde{m}}\neq0$ when $\tilde{m}$ is even.\\
\indent But this is the expansion for ${\mathcal P}_{{\phi }(2N)}$, what we are interested in is the expansion for $\ln({\mathcal P}_{{\phi }(2N)})$:
\begin{eqnarray}
\ln({\mathcal P}_{{\phi }(2N)})&=&\ln\left(\sum_{\tilde{m}} \tilde{c}_{\tilde{m}}|\phi-\phi'|^{\tilde{m}}\right)\nonumber\\
&=&\ln \tilde{c}_{0}-\sum_{m=1}\frac{(-1)^m}{m}\left(\frac{\sum_{\tilde{m}=1} \tilde{c}_{\tilde{m}} |\phi-\phi'|^{\tilde{m}}}{\tilde{c}_{0}}\right)^m\nonumber\\
&=&\sum_{m=0}c_m|\phi-\phi'|^m,
\end{eqnarray}
from where we see $c_1, c_3, \ldots, c_{2N-3}$ equal to zero, if all $\tilde{c}_1, \tilde{c}_3, \ldots, \tilde{c}_{2N-3}$ equal to zero; and $c_2$ is not zero since $\tilde{c}_2$ and $\tilde{c}_0$ are not zero.
\end{document}